\documentclass[gmd, manuscript]{copernicus}

\usepackage{grffile} \usepackage{subcaption}
\usepackage{xspace}

\usepackage{siunitx}

\DeclareSIUnit\year{y}

\newcommand{\PO}{PO\textsubscript{4}\xspace}
\newcommand{\DOP}{DOP\xspace}

\DeclareMathOperator*{\argmin}{argmin}

\DeclareMathOperator{\T}{^{\mkern-1.5mu\mathsf{T}}}

\DeclareMathOperator{\diag}{diag}

\begin{document}

\nolinenumbers

\title{Optimization of Model Parameters, Uncertainty Quantification and Experimental Designs for a Global Marine Biogeochemical Model}
\Author[1]{Joscha}{Reimer}
\affil[1]{Kiel University\\
Department of Computer Science\\
Algorithmic Optimal Control - CO$_2$ Uptake of the Ocean\\
24098 Kiel\\
Germany}
\correspondence{Joscha Reimer (joscha.reimer@email.uni-kiel.de)}
\correspondence{Joscha Reimer (joscha.reimer.edu@web.de)}
\runningtitle{Optimization of Model Parameters, Uncertainty Quantification and Experimental Designs}
\runningauthor{J. Reimer}
\received{}
\pubdiscuss{} \revised{}
\accepted{}
\published{}
\firstpage{1}
\maketitle

\begin{abstract}
Methods for model parameter estimation, uncertainty quantification and experimental design are summarized in this paper. They are based on the generalized least squares estimator and different approximations of its covariance matrix using the first and second derivative of the model regarding its parameters.
The methods have been applied to a model for phosphate and dissolved organic phosphorus concentrations in the global ocean. As a result, model parameters have been determined which considerably improved the consistency of the model with measurement results.
The uncertainties regarding the estimated model parameters caused by uncertainties in the measurement results have been quantified as well as the uncertainties associated with the corresponding model output implied by the uncertainty in the model parameters. This allows to better assess the model parameters as well as the model output.         
Furthermore, it has been determined to what extent new measurements can reduce these uncertainties. For this, the information content of new measurements has been predicted depending on the measured process as well as the time and the location of the measurement. This is very useful for planning new measurements.
\end{abstract}

\section{Introduction}

Computer models are a primary tools in natural sciences and contain parameters which are usually insufficiently known (cf. \cite{McGuffie2005, Neelin2010}). These parameters are usually estimated using noisy measurement data (cf. \cite{Aster2013,Seber2003}). This noise implies uncertainty in the estimated parameters as well as in the corresponding model output. This uncertainty is often not quantified, which, on the contrary, is essential to correctly assess the model parameters and the model output.

In order to counter this, we are going to summarize some techniques to estimate unknown model parameters and to quantify and reduce associated uncertainties. The presented methods are suited for computational complex models. We are going to demonstrate this using a model describing the phosphate and dissolved organic phosphorus concentrations in the global ocean. Phosphate is a limiting nutrient for marine phytoplankton and therefore influences the growth of phytoplankton and the absorption of atmospheric CO$_2$ by the ocean (cf. \cite[Chapter 4]{Bigg2003}).

Only uncertainties resulting from measurements are subject of this article. Model errors, i.e., the discrepancies between models and their modeled processes, are not captured as well as numerical errors, i.e., discrepancies between mathematical models and their (discretized) implementations. 

The methods, including methods for parameter estimation, uncertainty quantification and experimental design, are presented in Section \ref{sec: methods}. The marine model as application example is introduced in Section \ref{sec: model}. The results obtained for this model are presented in Section \ref{sec: results}. Finally, a conclusion is drawn in Section \ref{sec: conclusion}.

 \section{Methods for Parameter Estimation, Uncertainty Quantification and Experimental Design}
\label{sec: methods}

The generalized least squares estimator, as a model parameter estimation method, is summarized in this section together with its statistical assumptions and properties. Based on this, methods to quantify the uncertainty in the model parameters estimate and its corresponding model output are presented. They are built on approximations of the covariance matrix of the estimator and resulting approximate confidence intervals. Finally, optimal experimental design methods, which allow to reduce the uncertainty by optimally planned additional measurements, are briefly introduced.

\subsection{Model Parameter Estimation}
\label{subsec: parameter estimation}

An estimate $\hat{\theta}_n$ of the model parameters is usually obtained as the minimizer of an objective function $\phi_n$:
\begin{equation} \label{eq: parameter estiamte}
    \hat{\theta}_n := \argmin_{\theta \in \Omega} \phi_n(\theta)
    ,
\end{equation}
where $\Omega$ is some set of feasible model parameters and $n$ is the number of measurements.

By far, the most commonly used estimate is the (ordinary) least squares estimate (cf. \cite[Section 2.1]{Seber2003}, \cite[Section 3.1]{Pronzato2013}, \cite[Section 4.3]{Smith2013} and \cite[Section 3.1]{Walter1997}) where the objective function is:
\begin{equation*}
    \phi_n^{OLS}(\theta)
    :=
\|y_n - f_n(\theta)\|_2^2.
\end{equation*}
Here, $y_n$ denotes the vector of the measurement results and $f_n(\theta)$ the vector of model outputs corresponding to the measurement points and depending on the model parameters $\theta$.

However, we will use the generalized least squares estimate (cf. \cite[Subsection 2.1.4]{Seber2003}) with the objective function:
\begin{equation} \label{eq: generalized least squares: objective}
    \phi_n(\theta)
    :=
(y_n - f_n(\theta))\T {\mathcal{C}_n}^{-1} (y_n - f_n(\theta))
    ,
\end{equation}
where $\mathcal{C}_n$ is some positive definite matrix. If $\mathcal{C}_n$ is the identity matrix, this equals the ordinary least squares estimate and thus can be interpreted as a generalization. If $\mathcal{C}_n$ is a diagonal matrix, this corresponds to weighted least squares estimates.

The estimator corresponding to the generalized least squares estimate is the random vector:
\begin{equation*}
    \Theta_n
    :=
    \argmin_{\theta \in \Omega}
(Y_n - f_n(\theta))\T {\mathcal{C}_n}^{-1} (Y_n - f_n(\theta))
    ,
\end{equation*}
where $Y_n$ is a random vector of which the measurement results $y_n$ are a realization.

The estimator has some appealing properties under certain regularity conditions (cf. \cite{Jennrich1969}, \cite{Amemiya1983}, \cite[Section 12.2]{Seber2003}, \cite[Subsection 3.3.3]{Walter1997} and \cite[Section 3.1]{Pronzato2013}): It is consistent, asymptotically unbiased, asymptotically normal and asymptotically efficient. Hence, the estimator $\Theta_n$ converges almost surely to the desired model parameters if the number of measurements $n$ goes to infinity, making it the most accurate estimator among all asymptotically unbiased estimators.

One of the regularity conditions is that the statistical model, which includes the model function $f$ and the measurement noise, are correctly specified. For the generalized least squares estimator, this means that some true model parameters $\theta^* \in \Omega$ exist with:
\begin{equation} \label{eq: Y: assumed distribution}
    Y_n \sim \mathcal{N}(f_n(\theta^*), \sigma^2 \mathcal{C}_n),
\end{equation}
where $\sigma$ is some positive scalar. This implies that the model can describe the modeled process with appropriate parameters and that the measurement noise is unbiased and normally distributed with covariance matrix $\sigma^2 \mathcal{C}_n$.

If the assumed statistical model in \eqref{eq: Y: assumed distribution} is correct, the generalized least squares estimator is the maximum likelihood estimator. The desired model parameters, that should be estimated, are then $\theta^*$ and the consistency means then almost surely convergence to $\theta^*$.

If the assumed statistical model is incorrect, the generalized least squares estimator $\Theta_n$ is the quasi maximum likelihood estimator, also known as pseudo maximum likelihood estimator, regarding the set of probability distributions:
\begin{equation*}
    \mathcal{P}_n
    :=
    \{\mathcal{N}(f_n(\theta), \sigma^2 \mathcal{C}_n) ~|~ \theta \in \Omega\}
    .
\end{equation*}

This estimator is still consistent, asymptotically unbiased and asymptotically normal under certain regularity conditions (cf. \cite{White1981} and \cite{White1982}). However, the estimator no longer has to be efficient. Consistency means in this case the almost sure convergence to some $\theta^* \in \Omega$ so that $\mathcal{N}(f_n(\theta^*), \sigma^2 \mathcal{C}_n) \in \mathcal{P}_n$ has minimal difference to the probability distribution of $Y_n$ among all probability distributions in $\mathcal{P}_n$.

The other regularity conditions vary slightly depending on the reference. They usually includes, that the model function $f$ is twice continuously differentiable and that $\Omega$ is closed and bounded. Furthermore, $\theta^*$ must be uniquely identifiable, implying the measurement points must be chosen such that the model output at this points differs sufficiently for the model parameters $\theta^*$ compared to other model parameters $\theta \in \Omega$.

It should be noted that at some references only the ordinary least squares estimator is considered. However, their statements can be extended to the generalized least squares estimator by considering:
\begin{equation*}
    \phi_n(\theta)
    =
\| \tilde{y}_n - \tilde{f}_n(\theta) \|_2^2
    ,
\end{equation*}
as an ordinary least squares estimation with $\tilde{y}_n := {\mathcal{C}_n}^{-0.5} y_n$ and $\tilde{f}_n(\theta) := {\mathcal{C}_n}^{-0.5} f_n(\theta)$ (cf. \cite[Subsection 2.1.4]{Seber2003}).

\subsection{Uncertainty in Parameter Estimation}
\label{subsec: parameter uncertainty}

The uncertainty in the estimated model parameters $\theta_n$ implied by the noise in the measurement results can be described by the probability distribution of the estimator $\Theta_n$. Hence, we derive different approximations of this probability distribution in this subsection and calculate from these approximate confidence intervals for the unknown model parameters $\theta^*$.

In order to approximate the probability distribution of $\Theta_n$, we assume that:
\begin{equation} \label{eq: P: general distribution}
    \Theta_n
\sim
    \mathcal{N} \left(\theta^*, \mathcal{V}_n \right)
    ,
\end{equation}
where $\mathcal{V}_n$ is the covariance matrix of $\Theta_n$. This is reasonable due to the asymptotically normal distribution and the asymptotically unbiasedness of the estimator $\Theta_n$. They are ensured if the previous mentioned regularity conditions are fulfilled regardless of whether the assumed statistical model \eqref{eq: Y: assumed distribution} is correct or not.

The error made due to assumption \eqref{eq: P: general distribution} is small, if $n$ is sufficiently large. If the model function $f_n$ is linear regarding the model parameters and the statistical model \eqref{eq: Y: assumed distribution} is correct, \eqref{eq: P: general distribution} is a consequence (cf. \cite[Section 7.2]{Smith2013} and \cite[Section 2.6]{Tenorio2017}). Thus, even if $n$ is low, the error made due to assumption \eqref{eq: P: general distribution} is small if the second and higher derivatives of $f_n$ are close to zero and the statistical model assumed in \eqref{eq: Y: assumed distribution} is sufficiently close to reality.

To approximate the covariance matrix $\mathcal{V}_n$, we use three different approximations. In order to introduce these, we first define $J_n(\hat{\theta}_n)$ as the Jacobian matrix of the model function $f_n$ and $H_n(\hat{\theta}_n)$ as the Hessian matrix of the objective function $\tfrac{1}{2} \phi_n$ both at the estimate $\hat{\theta}_n$ as well as:
\begin{equation}
    F_n(\hat{\theta}_n) := J_n(\hat{\theta}_n)\T \mathcal{C}_n^{-1} J_n(\hat{\theta}_n)
    .
\end{equation}
$F_n(\hat{\theta}_n)$, sometimes instead $H_n(\hat{\theta}_n)$, is called the Fisher information matrix for the model parameter $\hat{\theta}_n$ (cf. \cite[Section 3.10]{Pukelsheim2006}).

The three approximations of $\mathcal{V}_n$ are:
\begin{gather}
    \label{eq: P: covariance matrix: Jacobian}
    \mathcal{V}_n^{(F)}(\hat{\theta}_n)
    :=
\sigma^2
    F_n(\hat{\theta}_n)^{-1}
    ,
    \\
    \label{eq: P: covariance matrix: Hessian}
    \mathcal{V}_n^{(H)}(\hat{\theta}_n)
    :=
\sigma^2
    H_n(\hat{\theta}_n)^{-1}
    ,
    \\
    \label{eq: P: covariance matrix: Jacobian and Hessian}
    \mathcal{V}_n^{(F,H)}(\hat{\theta}_n)
    :=
\sigma^2
    H_n(\hat{\theta}_n)^{-1} F_n(\hat{\theta}_n) H_n(\hat{\theta}_n)^{-1}
    .
\end{gather}

$\mathcal{V}_n^{(F)}(\hat{\theta}_n)$ is the most common of these approximations (cf. \cite[Subsection 2.1.2]{Seber2003}, \cite[Section 7.3]{Smith2013}, \cite[Section 5.2]{Tenorio2017}, \cite[Subsection 5.3.1]{Walter1997} and \cite{Donaldson1987}). It is derived by assuming the correctness of the statistical model \eqref{eq: Y: assumed distribution} and applying the linear least squares theory (cf. \cite[Section 7.2]{Smith2013} and \cite[Section 2.6]{Tenorio2017}) to the linearized model $l_n(\theta) := f_n(\hat{\theta}_n) + J_n(\hat{\theta}_n) (\theta - \hat{\theta}_n)$.

$\mathcal{V}_n^{(H)}(\hat{\theta}_n)$ is justified by the asymptotic theory for nonlinear least squares estimation (cf. \cite[Subsection 12.2.3]{Seber2003}), where under the assumed regularity conditions, $\tfrac{1}{n} F_n(\hat{\theta}_n)$ equals $\tfrac{1}{n} H_n(\hat{\theta}_n)$ asymptotically if the statistical model \eqref{eq: Y: assumed distribution} is correct.

$\mathcal{V}_n^{(F,H)}(\hat{\theta}_n)$ is derived by the asymptotic theory of quasi maximum likelihood estimators (cf. \cite{White1982}) where it is not necessary that the assumed statistical model \eqref{eq: Y: assumed distribution} is correct.

If the statistical model \eqref{eq: Y: assumed distribution} is correct, all three approximations are asymptotically equal and approach asymptotically the true asymptotic covariance matrix of $\Theta_n$ (cf. \cite{White1982}). Nevertheless, they are usually not equal for a finite number of measurements, because:
\begin{equation*}
    H_n(\hat{\theta}_n)
    =
F_n(\hat{\theta}_n)
    +
\sum_{k = 1}^{n} H^f_{k}(\hat{\theta}_n) \left( \mathcal{C}_n^{-1} (y_n - f_n(\hat{\theta}_n)) \right)_{k}
    ,
\end{equation*}
where $H^f_{k}(\hat{\theta}_n)$ is the Hessian matrix of the model at the $k$-th measurement point with respect to its parameters evaluated at $\hat{\theta}_n$. However, if $f$ is a linear function, all three approximations are equal, since $H^f_{k}(\hat{\theta}_n) = 0$.

It is not obvious which of these three approximations entails the smallest error if the statistical model \eqref{eq: Y: assumed distribution} is correct and $f$ is nonlinear. Different recommendations can be found in the literature (cf. \cite{Donaldson1987}, \cite{Cao2012}, \cite{Cao2009} and \cite{Efron1978}).

However, if the statistical model \eqref{eq: Y: assumed distribution} is not correct, which is the common case, only $\mathcal{V}_n^{(F,H)}(\hat{\theta}_n)$ approaches asymptotically the true asymptotic covariance matrix of $\Theta_n$ (cf. \cite{White1982}) and, hence, should be preferred.

If $\sigma$ is unknown, it can be estimated by:
\begin{equation}
    \hat{\sigma}_n^2
    :=
\tfrac{1}{n}
    \phi(\hat{\theta}_n)
    .
\end{equation}
This is an estimation of a consistent and asymptotically efficient estimator for $\sigma^2$, if the assumed statistical model in \eqref{eq: Y: assumed distribution} is correct (cf. \cite[Subsection 2.2.1]{Seber2003}). Otherwise $\hat{\sigma}_n^2$ converges almost surely to $\sigma^2 + e$, where $e \geq 0$ is the prediction mean square error, (cf. \cite[Subsection 12.2.4]{Seber2003} and \cite[Theorem 1]{Pazman2006}). Hence, $\hat{\sigma}_n^2$ usually overestimates $\sigma^2$ in this case.

After the covariance matrix $\mathcal{V}_n$ is approximated by $\hat{\mathcal{V}}_n$, approximate confidence intervals for the unknown model parameters $\theta^*$ can be constructed. For this, we first note that \eqref{eq: P: general distribution} considered component by component implies:
\begin{equation*}
    (\Theta_n)_i
    \sim
    \mathcal{N} \left( (\theta^*)_i, (\mathcal{V}_n)_{ii}\right)
    ,
    \qquad
    \text{ for all }
    \enspace
    i \in \{1, \ldots, m\}
    ,
\end{equation*}
 where $m$ is the number of model parameters.
Thus a confidence interval $(\mathcal{I}_n)_i$ for the $i$-th unknown true model parameter $(\theta^*)_i$ with approximate confidence level $\gamma$ can be constructed as:
\begin{equation} \label{eq: confidence intervals for p with estimated sigma}
\begin{gathered}
    (\mathcal{I}_n)_i
    :=
    [(\hat{\theta}_n)_i - (\alpha_n)_i, (\hat{\theta}_n)_i + (\alpha_n)_i]
    ,
    \qquad
    \text{ with }
    \enspace
    (\alpha_n)_i
    :=
    q\left( \tfrac{1 + \gamma}{2}, n - m \right) \sqrt{(\hat{\mathcal{V}}_n)_{ii}} 
    ,
\end{gathered}
\end{equation}
(cf. \cite[Section 5.1]{Seber2003} and \cite[Section 7.3]{Smith2013}) where $q(\beta, k)$ denotes the $\beta$ percentile of the t-distribution with $k$ degrees of freedom. Typical values are listed in Table \ref{tab: typical values t percentiles}.
\begin{table}[H]
    \centering
    \begin{tabular}{l|lllll}
    $\gamma$~\textbackslash~$k$    & $10$  & $10^2$  & $10^3$  & $10^4$  & $10^5$ \\
        \hline 
    $90\%$  & 1.812 & 1.660 & 1.646 & 1.645 & 1.645\\
    $95\%$  & 2.228 & 1.984 & 1.962 & 1.960 & 1.960\\
    $98\%$  & 2.764 & 2.364 & 2.330 & 2.327 & 2.326\\    
    $99\%$  & 3.169 & 2.626 & 2.581 & 2.576 & 2.576\\
    \end{tabular}    
    \par\smallskip
    \caption{Typical values for $q \left( \tfrac{1 + \gamma}{2}, k \right)$ rounded to three decimal places.}
    \label{tab: typical values t percentiles}
\end{table}

The justification of \eqref{eq: confidence intervals for p with estimated sigma} is that:
\begin{equation*}
    \frac{(\Theta_n)_i - (\theta^*)_i}{\sqrt{(\mathcal{V}_n)_{ii}}}
    \sim
    t_{n - m}
    ,
\end{equation*}
(cf. \cite[Section 5.1]{Seber2003}) where $t_{n - m}$ is the t-distribution with $n -m$ degrees of freedom.

The advantage of the previously described approach to quantify the uncertainty is that it is calculable without much computational effort, provided that the associated derivatives can be evaluated, at least approximately, without too much effort.

Another option to quantify the uncertainty regarding the model parameters are Monte Carlo simulations (cf. \cite[Section 5.2]{Walter1997}). Here fictitious measurement data vectors are generated several times and each time the resulting model parameters are estimated. From these estimates, confidence intervals for the unknown true model parameters $p$ could be calculated as well as statistical properties, like the expected value or the covariance matrix, of the estimator $\Theta_n$.

The fictitious measurement data vectors can be generated using sampling methods like random sampling or Latin hypercube sampling as well as resampling methods like jackknife of bootstrap methods.

This Monte Carlo approach provides more accurate results than the previously described approximations if the number of fictitious measurement data vectors is large. However, the computational effort using this approach is enormous in comparison to the described above because a parameter estimation has to be performed several times. Hence, it is not applicable to our computational expensive model.

\subsection{Uncertainty in Model Output}
\label{subsec: model uncertainty}

The uncertainty in the model parameters implies an uncertainty in the model output. This can be quantified in the same ways as the uncertainty in the model parameters. First a probability distribution of the corresponding random vector and then confidence intervals are approximated.

The uncertainty can be considered on the whole model output or only at some points of interest. Let $\tilde{f}$ denote the function that maps the model parameters to the model output whose uncertainty should be quantified. This can be the hole model output or only a subset.

The probability distribution of $\tilde{f}(\Theta_n)$ can then be used to describe the uncertainty in the model output due to the uncertainty in the model parameters. It can be approximated by:
\begin{equation} \label{eq: f: distribution}
    \tilde{f}(\Theta_n)
    \sim
    \mathcal{N}(\tilde{f}(\theta^*), \mathcal{W}_{n}(\hat{\theta}_n))
    ,
\end{equation}
with
\begin{equation} \label{eq: f: covariance matrix}
    \mathcal{W}_{n}(\hat{\theta}_n)
    :=
    J_{\tilde{f}}(\hat{\theta}_n) \mathcal{V}_{n}(\hat{\theta}_n) J_{\tilde{f}}(\hat{\theta}_n)\T
    .
\end{equation}
where $J_{\tilde{f}}(\hat{\theta}_n)$ is the Jacobian matrix of $\tilde{f}$ evaluated at $\hat{\theta}_n$ and $\mathcal{V}_{n}(\hat{\theta}_n)$ is an approximation of the covariance matrix of $\Theta_n$.

The approximations \eqref{eq: f: distribution} and \eqref{eq: f: covariance matrix} can be derived by assuming $\Theta_n \sim \mathcal{N} \left(\theta^*, \mathcal{V}_{n}(\hat{\theta}_n) \right)$ and calculating the probability distribution of $\tilde{l}(\Theta_n)$ where $\tilde{l}$ is a linearization of $\tilde{f}$. Another justification is the delta method (cf. \cite[Theorem 2.27]{Tenorio2017}) which allows to calculate the asymptotic probability distribution of $\tilde{f}(\Theta_n)$ if the asymptotic probability distribution of $\Theta_n$ is known. 

Several approximations of the covariance matrix of $\Theta_n$, namely $\mathcal{V}_n^{(F,H)}(\hat{\theta}_n)$, $\mathcal{V}_n^{(F)}(\hat{\theta}_n)$ and $\mathcal{V}_n^{(H)}(\hat{\theta}_n)$, were introduced in the previous subsection. Define $\mathcal{W}_{n}^{(F,H)}(\hat{\theta}_n)$, $\mathcal{W}_{n}^{(F)}(\hat{\theta}_n)$ and $\mathcal{W}_{n}^{(H)}(\hat{\theta}_n)$ as described in \eqref{eq: f: covariance matrix} using $\mathcal{V}_n^{(F,H)}(\hat{\theta}_n)$, $\mathcal{V}_n^{(F)}(\hat{\theta}_n)$ and $\mathcal{V}_n^{(H)}(\hat{\theta}_n)$, respectively. $\mathcal{W}_n^{(F,H)}(\hat{\theta}_n)$ is a good choice if the assumed statistical model might be incorrect. $\mathcal{W}_n^{(F)}(\hat{\theta}_n)$ and $\mathcal{W}_n^{(H)}(\hat{\theta}_n)$ are also reasonable if the assumed statistical model \eqref{eq: Y: assumed distribution} is correct.

Looking at a single point of interest, \eqref{eq: f: distribution} implies:
\begin{equation}
    (\tilde{f}(\Theta_n))_i
    \sim
    \mathcal{N} \left( (\tilde{f}(\theta^*))_i, (\mathcal{W}_{n}(\hat{\theta}_n))_{ii} \right)
    .
\end{equation}

Thus a confidence interval $\tilde{\mathcal{I}}_i$ for $(\tilde{f}(\Theta_n))_i$ with approximate confidence level $\gamma$ can be constructed, in the same way as in the previous subsection, as:
\begin{equation} \label{eq: confidence intervals for f with estimated sigma}
\begin{gathered}
    (\tilde{\mathcal{I}}_n)_i
    :=
    [(\tilde{f}(\hat{\theta}_n))_i - (\tilde{\alpha}_n)_i, (\tilde{f}(\hat{\theta}_n))_i + (\tilde{\alpha}_n)_i]
    \qquad
    \text{ with }
    \enspace
    (\tilde{\alpha}_n)_i
    :=
    q\left( \tfrac{1 + \gamma}{2}, n - m \right) \sqrt{(\mathcal{W}_n)_{ii}} 
    .
\end{gathered}
\end{equation}

Again, the advantage of these approximation is that they are calculable without much computational effort.

Instead of this approximation, Monte Carlo simulations could again be used to quantify the uncertainty. This time several model parameter vectors have to be generated from the assumed probability distribution of $\Theta_n$. For each of this model parameter vectors, the model output at the points of interest have to be evaluated. From these model evaluations, confidence intervals could be calculated as well as statistical properties of $\tilde{f}(\Theta_n)$, like its expected value or covariance matrix.

Again, this approach provides more accurate results than the approximation \eqref{eq: f: distribution} and \eqref{eq: f: covariance matrix} if the number of generated model parameter vectors is large but the computational effort is extensive compared to these approximations. Hence, it is not applicable here.

\subsection{Uncertainty Reduction using Optimal Experimental Design Methods}
\label{subsec: uncertainty reduction}

The uncertainty in the model parameters as well as the model output can be reduced by additional measurements. However, not all measurements reduce the uncertainty equally. The idea of optimal experimental design methods (cf. \cite{Pukelsheim2006}, \cite[Chapter 6]{Walter1997} and \cite[Subsection 5.13]{Seber2003}) is to design the measurements such that the resulting uncertainty is minimized and, hence, the information gain is maximized.

The design of a measurement includes everything that characterizes the measurement, involving the place and time of the measurement. If several different processes are modeled, it also includes which process is measured. Furthermore, multiple measuring techniques might be choosable which might result in different measurement accuracies.

One of the key observations for optimal experimental design methods is that for a given $\hat{\theta}_n$, the actual measurement results are not needed for the calculation of $\mathcal{V}_n^{(F)}(\hat{\theta}_n)$.  Hence, it can also be calculated including planned measurements that have not yet been carried out. The same applies to $\mathcal{W}_n^{(F)}(\hat{\theta}_n)$. Thus the new uncertainty resulting from additional measurements can be predicted with these values.

Using $\mathcal{V}_n^{(F)}(\hat{\theta}_n)$ and $\mathcal{W}_n^{(F)}(\hat{\theta}_n)$ is justified if the assumed statistical model \eqref{eq: Y: assumed distribution} is correct. Otherwise they may not be consistent estimations of the corresponding covariance matrices. Nevertheless, they can be used under certain regularity conditions (cf. \cite{Pazman2006}) to assess measurement designs. $\mathcal{V}_n^{(H)}(\hat{\theta}_n)$ and $\mathcal{V}_n^{(F, H)}(\hat{\theta}_n)$ as well as $\mathcal{W}_n^{(F)}(\hat{\theta}_n)$ and $\mathcal{W}_n^{(F, H)}(\hat{\theta}_n)$ can not be used to predict the uncertainty reduction because they depend on the measurement results.

To compare the uncertainty or the information gain resulting from different measurement designs criteria (cf. \cite[Chapter 5]{Pukelsheim2006}, \cite[Section 6.1]{Walter1997} and \cite[Chapter 5]{Pronzato2013}) are established. These criteria quantify the uncertainty with a single value by mapping covariance matrices to scalar values. Typical design criteria are the sum of the diagonal values, the determinant and the maximal eigenvalue (cf. \cite[Chapter 6]{Pukelsheim2006}, \cite[Section 6.1]{Walter1997} and \cite[Subsection 5.1.2]{Pronzato2013}). The lower the values of these criteria are, the stronger the measurements would reduce the uncertainty.

The choice of an appropriate design criterion depends on the purpose of the additional measurements. In particular, whether the uncertainty in the model parameters or in the model outputs should be reduced and how much emphasis is placed on the reduction of individual model parameters or model outputs.

We have used two different design criteria. They are easy to calculate and to interpret. The first one aims at reducing uncertainty in the model parameters itself and is defined as:
\begin{equation}
\label{eq: oed criterion: relative average parameter uncertainty}
    \psi(\mathcal{V}_n^{(F)}(\hat{\theta}_n), \hat{\theta}_n)
    :=
    \frac{1}{m} \sum\limits_{k = 1}^{m} \frac{\sqrt{(\mathcal{V}_n^{(F)}(\hat{\theta}_n))_{ii}}}{(\hat{\theta}_n)_i}
    .
\end{equation}
This is the average of the relative uncertainty in each model parameter, quantified by the standard deviation of the corresponding estimator divided by the parameter value. Designs are therefore preferred which evenly reduce the uncertainty in each of the model parameters. The average of the absolute uncertainties, i.e., the average of the uncertainties not divided by the parameter values, is less useful, if typical model parameters are of different orders of magnitude.

The second design criteria is:
\begin{equation}
\label{eq: oed criterion: relative average model output uncertainty}
    \psi_{\tilde{f}}(\mathcal{W}_n^{(F)}(\hat{\theta}_n), \hat{\theta}_n)
    :=
\frac{1}{l}
    \sum\limits_{k = 1}^{l}
    \left( \sum\limits_{i \in I_k} \sqrt{(\mathcal{W}_n^{(F)}(\hat{\theta}_n))_{ii}} \right)
    \left( \sum\limits_{i \in I_k} \tilde{f}_i(\hat{\theta}_n) \right)^{-1}
\end{equation}
where $I_k$ is the set of indices corresponding to the output of the $k$-th modeled process of the numbered $l$ modeled processes. This criterion prefers designs which reduce the uncertainty in the model output at the points of interest evenly over all modeled processes.

Again, the absolute uncertainty might be less useful if the typical total model output for each process and thus its typical total absolute uncertainty differs by several orders of magnitude. The uncertainty relative to each individual model output is not useful either if some model outputs are zero or close to zero. 

It should be straight forward to modify the criteria to the specific purpose of the additional measurements or to formulate new ones specially suited. Designs that minimize the criterion among all feasible designs are called (local) optimal designs. Local refers to the dependency on the parameter estimate $\hat{\theta}_n$.

The information gain by additional measurements can be quantified by subtracting the value of the criteria using only the previous designs with the value of the criteria using the previous and the additional designs.

Sometimes, it might be useful to assign a cost value to each design that quantifies the financial cost or the time effort associated with this measurement, so the predicted information gains relative to their costs can be considered. This allows to define optimal designs in relation to their costs or to choose designs up to a certain cost limit.

After carrying out the chosen additional measurements, their measurement results should be used together with the previous measurement results to update the estimate of the model parameters. Using this new estimate new measurements could be designed. This allows to include the information in the previous measurements in the planning of the next measurements. This iterative process is called sequential optimal experimental design (cf. \cite[Subsection 6.4.2]{Walter1997} and \cite[Subsection 5.13.3]{Seber2003}) and is particularly suitable if new measurements have to be planned repeatedly.

\subsection{Computational Details}
\label{subsec: computational details}

Several computational details regarding the estimation of the model parameters, as described in Subsection \ref{subsec: parameter estimation}, are summarized in the following subsection.

\subsubsection*{Optimization Algorithm}
\label{sec: optimization algorithm}

A number of optimization algorithms exist which can be used to calculate the model parameter estimate $\hat{\theta}_n$ by minimizing the objective function $\phi_n$. They can basically be divided into two categories: derivative based (cf. \cite{Gill1981} and \cite{Nocedal2006}) and derivative free algorithms (cf. \cite{Conn2009} and \cite{Rios2013}).

Usually derivative based optimization algorithms need fewer function evaluations to find a local minimum compared to derivative free optimization algorithms. However, they usually have more difficulties finding a global minimum. We try to take advantage of the rapid convergence of the derivative based optimization algorithm SQP discussed in \cite[Chapter 18]{Nocedal2006} and try to avoid its difficulty with finding global minimum by combining it with the globalization algorithm OQNLP introduced in \cite{Ugray2007}.

This OQNLP algorithm finds the minimizer by starting multiple local minimizations from promising start points. To generate start points, a scatter-search algorithm similar to that described in \cite{Glover1998} is used. Thereafter, iteratively, local minima are searched by a local optimization algorithm from one of the start points. After each search, unpromising start points are removed from the set of start points by considering their value of the objective function and their distance to already found local minima. The algorithm terminates if all start points are used or removed. The local minimum with the lowest objective value is then identified as global minimum. This OQNLP algorithm is implemented in MATLAB (cf. \cite{Matlab2015a}) as  GlobalSearch algorithm in the Global Optimization Toolbox (cf. \cite[Chapter 3]{MatlabGlobalOptimizationToolbox2015a}).

This SQP algorithm iteratively solves the Karush–Kuhn–Tucker (KKT) equations, introduced in \cite{Karush1939} and \cite{Kuhn1951}, of the constrained optimization problem. For this purpose, a constrained quadratic subproblem is solved in each iteration using an active set strategy like described in \cite{Gill1981,Gill1991}. The solution of the quadratic subproblem is used as search direction for a line search procedure similar to that described in \cite{Han1977}, \cite{Powell1978a} and \cite{Powell1978}. The quadratic subproblem is formulated using the value of the objective function and its first derivative as well as an approximation of its second derivative. The BFGS method, developed by \cite{Broyden1970}, \cite{Fletcher1970}, \cite{Goldfarb1970} and \cite{Shanno1970}, is used as a quasi-Newton update for this approximation together with an correction technique, described in \cite{Powell1978a}, which keeps the approximated Hessian positive define. The SQP algorithm is implemented in MATLAB as fmincon algorithm in the Optimization Toolbox (cf. \cite[Chapter 6]{MatlabOptimizationToolbox2015a}).

\subsubsection*{Scaling of the Objective Function}

Many optimization algorithms, like the one we have used, are not invariant to scaling therefore it is essential to scale the objective function (cf. \cite[Section 7.5 and 8.7]{Gill1981}, \cite[Section 7.3]{Smith2013}, \cite[Section 2.2]{Nocedal2006} and \cite[Section 7.1]{Dennis1996}) for a fast and accurate determination of a minimum.

Hence, for the estimation of the model parameters, the model parameters in the objective function and the objective function values are scaled, as described in \cite[Section 7.5 and 8.7]{Gill1981}. The scaled parameters typically range from -1 to 1 and the objective function values typically be around 1.

\subsubsection*{Evaluating of the Objective Function}

The objective function of the generalized least squares estimator \eqref{eq: generalized least squares: objective} can be evaluated in many different ways. In the following, we describe a fast and numerically accurate way.

The objective function value that should be evaluated is:
\begin{equation*}
    \phi(p) = (y - f(p))\T \mathcal{C}^{-1} (y - f(p))
    .
\end{equation*}
We have omitted the index $n$ for the sake of simplicity. Define $S := \diag(\mathcal{C})^{0.5}$
to be the diagonal matrix containing the square root of the diagonal values of $\mathcal{C}$ and $B := S^{-1} A S^{-1}$.

Decompose $B$ by $LDL\T$ representation, meaning a lower triangle matrix $L$ with ones on the diagonal and diagonal matrices $D$ with positive values. It is important to notice that this has to be done only once and not for every evaluation of the objective function.

The objective function $\phi$ is then evaluated by first evaluating:
\begin{equation*}
    \psi(p)
    :=
    D^{-0.5} L^{-1} S^{-1} (y - f(p))
    ,
\end{equation*}
from right to left, where instead of the inverse of $L$ the corresponding linear equation is solved using forward substitution. Then objective function value is evaluated by:
\begin{equation*}
    \phi(p)
    =
    \psi(p)\T \psi(p)
    .
\end{equation*}

 \section{A Marine Phosphorus Cycle as Application Example}
\label{sec: model}

We use a model for the phosphate and dissolved organic phosphorus concentrations in the global ocean as application example for the parameter estimation, uncertainty quantification and experimental design methods described in Section \ref{sec: methods}.

First, the used circulation model and the biogeochemical model are introduced in this section. Then, the model parameters are described together with different guesses of their values. Next, the calculation of an annual periodic state is explained as well as a fast way to calculate the derivative of the model output regarding the model parameters. Finally, the measurement data used for parameter estimation are described.

\subsection{Circulation Model}
\label{subsec:transport model}

We have used the Transport Matrix Method (TMM) introduced in \cite{Khatiwala2005} to simulate the advection and diffusion of passive tracers in the ocean (cf. \cite{Khatiwala2007}). This method has already been used in various studies (cf. \cite{Weber2010, Kriest2010, Kriest2012, Priess2013, Graven2012}).

The TMM utilizes that the continuous advection-diffusion equation:
\begin{equation} \label{equ: continuous advection-diffusion}
    \frac{\partial Y_i}{\partial t} = \underbrace{\nabla \cdot (K \nabla \cdot Y_i))}_{\text{diffusion}} - \underbrace{\nabla \cdot (V Y_i)}_{\text{advection}} + \underbrace{S_i(Y_1, \ldots, Y_m, \theta)}_{\text{sources and sinks}} \text{~for~} i \in 1, \ldots, m
    ,
\end{equation}
where $Y_1, \ldots, Y_m$ denote the concentrations of the $m$ tracers, $K$ the diffusion coefficient, $V$ the velocity and $S_1, \ldots, S_m$ the source-and-sink-terms depending on the model parameters $\theta$, can be written in the discretized form as a matrix equation:
\begin{equation} \label{equ: discrete advection-diffusion}
    y^{(n+1)} = A_i^{(n)} (A_e^{(n)} y^{(n)} + s^{(n)}(\theta) \Delta t)
    .
\end{equation}
$y^{(n)}$ denotes the vector of all tracer concentrations at all grid points of the circulation model in the discretized form at time step $n$, $s^{(n)}(\theta)$ the discretized version of the source-and-sink-terms depending on the model parameters $\theta$ and $\Delta t$ the time step in the discretization. The matrices $A_i^{(n)}$ and $A_e^{(n)}$, called transport matrices, result from the discretization of the advection and diffusion terms where $A_i^{(n)}$ belongs to the implicit part and $A_e^{(n)}$ to the explicit part of the discretization.

The approach of the TMM is to determine the elements of these matrices by utilizing a general circulation model. For this, the general circulation model is executed several times with different suitable chosen tracer concentrations.

We use monthly averaged transport matrices (cf. \cite{Khatiwala2005, Khatiwala2007}), calculated with the MIT general circulation model (cf. \cite{Marshall1997b, Marshall1997a, Marshall1998}). At the middle of each month the corresponding transport matrix has been used. Elsewhere a linear interpolation of the two transport matrix closest to the point in time were used.

A spatial resolution of 2.8125 degree and 15 vertical layers with increasing depths was used at the construction of the transport matrices. Hence, this is also the resolution of our circulation model. The resolution corresponds to 64 boxes in north-south direction and 128 boxes in west-east direction.

For the temporal resolution, $\Delta t = 2880^{-1}$ \si{\year} has been chosen which corresponds to a time step of roughly three hours. Hence, daytime dependent processes can be resolved.

\subsection{Biogeochemical Model} \label{subsec: biogeochemical model}

The biogeochemical model contains phosphate (\PO) and dissolved organic phosphorus (\DOP) and is part of the ocean carbon model, described in \cite{Dutkiewicz2005}, of the MIT Integrated Global System Model Version 2 (IGSM2), described in \cite{Sokolov2005}. This model and some variants are used frequently(cf. \cite{Parekh2005, Parekh2006, Najjar1998, Najjar2007, Kwon2009, Kriest2010, Kriest2012, Priess2013}). It is briefly described in the following where we stick to the notation in \cite{Dutkiewicz2005}.

The concentration of \PO and \DOP at layer $i$ are described by the following source-minus-sink terms:
\begin{align}
    S_{\PO}(i)
    & =
    -J_{prod}(i) + \kappa_{re} DOP(i) + \Delta F(i)
    ,
    \\
    S_{\DOP}(i)
    & =
    f_{DOP} J_{prod}(i) - \kappa_{re} DOP(i)
    .
\end{align}
Here, $J_{prod}$ denote the biological production (net community productivity). A fraction $f_{DOP}$ of this biological production remains suspended as \DOP. The remainder $(1-f_{DOP})$ becomes particulate organic phosphorus (POP) which sinks to depths and instantly remineralizes to \PO which is modeled by $\Delta F(i)$. The \DOP remineralizes back to \PO with rate $\kappa_{re}$. $f_{DOP}$ and $\kappa_{re}$ are model parameters.

The biological production:
\begin{equation}
    J_{prod}(i) := \alpha \frac{PO_4(i)}{PO_4(i) + \kappa_{PO_4}} \frac{I(i)}{I(i) + \kappa_{I}}
    ,
\end{equation}
is modeled by Michaelis-Menten kinetics depending on the available light $I$ and the nutrient \PO similar to \cite{McKinley2004}. The corresponding half saturation constants $\kappa_{\PO}$ and $\kappa_{I}$ are model parameters as well as the maximum community production rate $\alpha$.

The available light:
\begin{equation}
    I(i) := f_{PAR} Q_{SW} e^{-k z_c(i)}
    ,
\end{equation}
is modeled, as that portion of the short wave radiation $Q_{SW}$ that is photo-synthetically available and has not been attenuated by water. The short wave radiation $Q_{SW}$ is calculated by the atmosphere component of the IGSM2 as a function of time, latitude and ice cover (cf. \cite{Paltridge1976} and \cite{Brock1981}). The light attenuation coefficient of water $k$ is treated as model parameter.

The fraction of photo-synthetically available radiation is described by $f_{PAR}$. It only enters into the biological production $J_{prod}$ where only the ratio $\frac{\kappa_{I}}{f_{PAR}}$ is relevant. Due to this linear dependence, $f_{PAR}$ and $\kappa_{I}$ would not be uniquely identifiable if $f_{PAR}$ would be a model parameter as well. For this reason, $f_{PAR}$ is set constant to $f_{PAR} := 0.4$. This values is also used in the IGSM2.

Let $n$ be the numbers of layers. For each layer $i$, let $z_t(i)$, $z_c(i)$ and $z_b(i)$ be its top, centered and bottom depth, respectively and $\Delta z(i) := z_b(i) - z_t(i)$ its thickness.

The portion of the biological production which is exported as POP from layer $i$ to deeper layers is denoted by:
\begin{equation}
    E(i) := (1 - f_{DOP}) J_{prod}(i) \Delta z(i)
    .
\end{equation}

It is assumed that the sinking speed increases with depth following a power law relationship (cf. \cite{Najjar1998}) and that the exported POP instantly remineralizes to \PO. The flux $F(i)$ into layer $i \geq 2$ is then modeled as follows, where $i_e$ is the last layer in the euphotic zone:
\begin{equation}
    F(i) := \sum\limits_{j=1}^{\min(i_e, i-1)} E(i) \left( \frac{z_b(i-1)}{z_b(j)} \right)^{-a_{re}}
    .
\end{equation}
The change of the \PO concentration in layer $1 < i < n$ due to the flux is then:
\begin{equation}
    \Delta F(i) := 
    \sum\limits_{j=1}^{\min(i_e, i-1)} E(i) \left( \left( \frac{z_b(i-1)}{z_b(j)} \right)^{-a_{re}} - \left( \frac{z_b(i)}{z_b(j)} \right)^{-a_{re}} \right) (\Delta z(i))^{-1}
    . 
\end{equation}
It is also assumed that no POP is lost to the sediment. This means, all POP that enters the deepest box is instant remineralized:
\begin{equation}
    \Delta F(n) := 
    \sum\limits_{j=1}^{\min(i_e, i-1)} E(i) \left( \frac{z_b(i-1)}{z_b(j)} \right)^{-a_{re}} (\Delta z(i))^{-1}
    .
\end{equation}
In the topmost layer no \PO arise from sunk and remineralized POP:
\begin{equation}
    \Delta F(1) := 0
    .
\end{equation}

\subsection{Model Parameters} \label{subsec: model parameters}

The seven parameters of the biogeochemical model, described in Subsection \ref{subsec: biogeochemical model}, are considered as unknown model parameters. Furthermore, the global average phosphorus concentration, which is used to spin-up the model into annual periodic concentrations as described in Section \ref{subsec: spin-up}, is considered as an unknown model parameter as well. All these model parameters are listed in Table \ref{tab: model parameters: description} and their values shall be estimated.

\begin{table}[H]
    \centering
\begin{tabular}{l|l|l}
        Parameter         & Description                               & Unit \\
        \hline
        $\kappa_{re}$  & remineralization rate of \DOP             & \si{\per\year} \\
        $\alpha$          & maximum community production rate         & \si{\milli\mol\per\cubic\metre\per\year} \\
        $f_{DOP}$         & fraction new production going to \DOP     & - \\
        $\kappa_{\PO}$    & half saturation constant of \PO           & \si{\milli\mol\per\cubic\metre} \\
        $\kappa_{I}$      & half saturation constant of light         & \si{\watt\per\square\metre} \\
        $k$               & light attenuation coefficient of water    & \si{\per\metre} \\
        $a_{re}$       & power law remineralization coefficient    & - \\
        $p$               & average phosphorus concentration   & \si{\milli\mol\per\cubic\metre} \\
    \end{tabular}
\par\smallskip
    \caption{Parameters of the marine phosphorus cycle model.}
    \label{tab: model parameters: description}
\end{table}

Our initial guesses and bounds for the unknown model parameters are summarized in Table \ref{tab: model parameters: bounds and init values}. They are based on values used in other publications which are outlined next.
\begin{table}[H]
    \centering
    \begin{tabular}{l|llllllll}
                        & $\kappa_{re}$  & $\alpha$  & $f_{DOP}$ & $\kappa_{\PO}$    & $\kappa_{I}$  & $k$    & $a_{re}$  & $p$\\
        \hline
        initial guess   & 0.5               & 2         & 0.67      & 0.5               & 30            & 0.02   & 0.86        & 2.17\\
        lower bound     & 0.05              & 0.2       & 0.05      & 0.01              & 5             & 0.001  & 0.5          & 0.4\\
        upper bound     & 10                & 20        & 0.95      & 10                & 200           & 0.2    & 2            & 10\\
    \end{tabular}
    \par\smallskip
    \caption{Bounds and initial guesses for model parameters.}
    \label{tab: model parameters: bounds and init values}
\end{table}

For $\kappa_{re}$, 2 \si{\per\year} was used in \cite{Najjar1998} and in \cite{Dutkiewicz2005} based on \cite{Najjar1998}. 0.5 \si{\per\year} was used in \cite{Parekh2005} and in \cite{Kriest2010} based on \cite{Parekh2005}. In \cite{Najjar1998} different studies are summarized which had suggested that $\kappa_{re} \in [\frac{10}{7}, 5]$ \si{\per\year}.

3 \si{\milli\mol\per\cubic\metre\per\year} was used in \cite{Dutkiewicz2005} for $\alpha$, 2 \si{\milli\mol\per\cubic\metre\per\year} in \cite{Kriest2010} and 6 \si{\milli\mol\per\cubic\metre\per\year} in \cite{Parekh2005}. In \cite{McKinley2004}, different values were used for different ocean regions with 2.5 \si{\milli\mol\per\cubic\metre\per\year} as average value.

$f_{DOP} \in [0, 1]$ by definition of $f_{DOP}$. 0.67 was used in \cite{Dutkiewicz2005}, \cite{Kriest2010}, \cite{Najjar1998} and \cite{Parekh2005} all based on \cite{Yamanaka1997}. 0.7 was suggested in \cite{Platt1989}. Different studies are cited in \cite{Najjar1998} which had estimated $f_{DOP}$ in $[0.58, 0.77]$, $[0.65, 0.95]$, $[0.6, 0.7]$ or $[0.4, 0.8]$.

For $\kappa_{\PO}$, 0.5 \si{\milli\mol\per\cubic\metre} was used in \cite{Dutkiewicz2005} and \cite{Kriest2010} and $0.01$ \si{\milli\mol\per\cubic\metre} in \cite{McKinley2004}.

25 \si{\watt\per\square\metre} was used in \cite{Dutkiewicz2005} for $\kappa_{I}$ and 30 \si{\watt\per\square\metre} in \cite{Dutkiewicz2001}, \cite{Kriest2010}, \cite{McKinley2004} and \cite{Parekh2005}. In \cite{Dutkiewicz2001}, it was stated that $\kappa_{I}$ varies from 5 \si{\watt\per\square\metre} to 100 \si{\watt\per\square\metre} for different species of phytoplankton based on several cited studies.

For $k$, 0.02 \si{\per\metre} was used in \cite{Dutkiewicz2005} and \cite{Kriest2010}.

0.9 was used in \cite{Dutkiewicz2005} for $a_{re}$ based on \cite{Yamanaka1997} and \cite{Sarmiento1990}. 0.858 was used in \cite{Martin1987} and in \cite{Kriest2010} based on \cite{Martin1987}. Since the choice of $a_{re}$ is closely related to $z_b(i_e)$, the depth of the euphotic zone, its common values are presented as well. 100 \si{\metre} was chosen in \cite{Yamanaka1997}, \cite{Martin1987} and \cite{Maier-Reimer1993} and 75 \si{\metre} in \cite{Najjar1998}. 120 \si{\metre} was used in \cite{Kriest2010} and 130 \si{\metre} in \cite{McKinley2004}. We selected 120 \si{\metre} as well.

2.1701 \si{\milli\mol\per\cubic\metre} was used in \cite{Kriest2010} for $p$. 2.17 \si{\milli\mol\per\cubic\metre} is also the average phosphorus concentration of the climatological data provided by the World Ocean Atlas 2013 \cite{Garcia2014} and \cite{Reimer2019a} which are both based on the data of the World Ocean Database 2013 introduced in \cite{Boyer2013}.

\subsection{Simulation and Spin-up} \label{subsec: spin-up}

The previously described model has been simulated using the simulation package (\cite{simulation-0.4}) which is based on Python (\cite{Python-3.7}), NumPy (\cite{NumPy-1.17.3}), SciPy (\cite{Jones2019b} and \cite{Virtanen2019}), Matplotlib (\cite{matplotlib-3.1.1} and \cite{Hunter2007}), utillib (\cite{utillib-0.3}), the measurements software package (\cite{measurements-0.3}) and the matrix-decomposition library (\cite{matrix-decomposition-1.2} and \cite{Reimer2019}). The simulation package also includes many pre- and post-processing functions. For the actual parallelized evaluation of the model, it uses the simulation framework METOS3D (\cite{Piwonski2016}, \cite{Piwonski2013}) which is based on PETSc (Portable, Extensible Toolkit for Scientific Computation) (\cite{Balay2019}, \cite{Balay2019a}).

For each model simulation, the model has been spun up from constant concentrations to annual periodic concentrations. These constant concentrations were chosen so that the average phosphorus concentration $p$ was achieved. To check if an annual periodicity is reached, the concentrations at the beginning of two consecutive model years were compared. If these are equal, a periodic state is reached.

Usually, it took 5000 to 7500 model years until roughly annual periodic concentrations were achieved. Sometimes even more model years were needed. We used at most 10.000 model years. Thereafter, the average difference between concentrations at two consecutive model years was around $10^{-7}$. 

A model simulation with a spin-up of 10.000 model years has taken about four hours on four connected computer nodes with sixteen cores each and a clock rate of 2.1 GHz, respectively.

\subsection{Derivative}

Besides the model output itself, the derivatives of the model output regarding the model parameters are needed for the estimation of the model parameters as well as the uncertainty quantification and the design of additional measurements. We have approximated them using finite difference quotients (cf. \cite[Section 25.3]{Abramowitz1972}, \cite[Section 4.2]{Dennis1996}, \cite[Section 8.1]{Nocedal2006} and \cite[Section 8.6]{Gill1981}). For this, appropriate finite difference quotients and step sizes must be selected.

Central finite difference quotients ( \cite[Equation 25.3.21]{Abramowitz1972} and \cite[Subsubsection 8.6.1.2]{Gill1981}) have been used for the first order partial derivatives. They have a second order approximation error and are, thus, very accurate with an appropriate step size. For the second order partial derivatives, finite difference quotients \cite[Equation 25.3.23 and Equation 25.3.27]{Abramowitz1972}, with a second order approximation error as well, have been used.

Two additional function evaluations are needed for each approximation of the first order partial derivative. If the same step size is used for approximating the second order partial derivatives, two more function evaluations are needed for the second order partial derivative regarding two different variables and no additional function evaluations are needed for the second order partial derivative regarding one variable.

To reduce the number of additional function evaluations, finite difference quotients with first order approximation error (cf. \cite[Section 4.2]{Dennis1996} and \cite[Section 8.6]{Gill1981}), like forward or backward finite difference quotients for the first order partial derivatives, could be used. However, in our application example, the additional function evaluations correspond only to a small part of the total computational effort, as explained below. For that reason, we use the more accurate finite difference quotients described in the previous paragraphs.

The choice of the step size in the finite difference quotients is always a compromise between a small error in replacing the derivative by the finite difference quotient and a small error in the floating point arithmetic. Recommended step sizes are usually a constant, depending on the used finite difference quotient, multiplied by the typical magnitude of the model parameter (cf. \cite[Section 5.6]{Dennis1996} and \cite[Section 8.1]{Nocedal2006}). These constants are the third and the fourth root of the machine precision, for the first order and second order finite difference quotient, respectively (cf. \cite[Subsection 8.6.1]{Gill1981}). These are roughly $10^{-5}$ and $10^{-4}$ for 64 bit floating point numbers.

Larger and smaller step sizes were also tested. However, too strong deviation from the recommended step size, usually by more than two orders of magnitude, result in unrealistic values.

To evaluate the finite difference quotients, we have first spun up the model with the unchanged model parameters. Usually this spin-up is needed anyway. The spin-ups for the slightly changed model parameters in the finite difference quotients were then started with the annual periodic concentrations, obtained from the spin-up with the unchanged model parameters, instead of the usually used constant concentrations. For the derivative regarding the average phosphorus concentration, the annual periodic concentration were slightly modified to match the average phosphorus concentration.

Using the annual periodic concentrations from the spin-up with the unchanged model parameters accelerates the evaluation of the derivative significantly because much fewer model years are needed to achieve annual periodic concentrations for the slightly changed model parameters. Tests with different model parameters have shown that usually only a few hundred model years are needed. Thus, we have used at most 500 years for the spin-ups for the slightly changed model parameters. Hence, the complete evaluation of the first derivative with central finite differences needs at most $80\%$ more computational effort than the evaluation of the model itself.

\subsection{Measurement Data}
\label{subsec: data}

We used the measurement data for phosphate provided by the World Ocean Database 2013, presented in \cite{Boyer2013} and \cite{WOD2013UsersManual}, for the model parameter estimation. We limited ourselves to the data that have passed all quality checks \cite[Section 3]{WOD2013UsersManual} and where the measurement points are inside the computational domain. These were about 2.2 million measurements.

For dissolved organic phosphorus, generally far less measurement data were available. We used almost 400 measurements obtained from \cite{Landolfi2005}, \cite{Landolfi2008} and \cite{Yoshimura2007}. These data were quality checked as well and implausible data were removed together with data outside of the computational domain.

The corresponding standard deviations and the correlation matrix were estimated as described in \cite{Reimer2019a} using the spatial resolution described in Subsection \ref{subsec:transport model} and a monthly temporal resolution.

Here, the standard deviation in each space-time grid box was estimated using the sample standard deviation in each grid box where at least four values are available. Otherwise the standard deviation was interpolated for phosphate. For dissolved organic phosphorus, the average of its estimated standard deviations was used, since too few data are available for a meaningful interpolation. Furthermore, we used 0.1 as a lower bound for the standard deviations. This corresponds to the usual accuracy of the measurement data and prevents a disproportional weighting of measurement results with a very small sample standard deviation.

The correlation between different space-time grid boxes was estimated using the sample correlation where at least thirty-five value pairs were available. Otherwise the correlation is assumed to be zero. From these individual estimates, a valid correlation matrix was calculated using the algorithm described in \cite{Reimer2019}.

The objective of the algorithm is to find a valid correlation matrix which is close to the original matrix and has a low condition number. A low condition number is important because otherwise small inaccuracies by numerical methods or measurements, are amplified and could dominate the evaluation of the objective function. The algorithm has a parameter which controls the weighting between a small difference to the original matrix and a small condition number. We have chosen 0.1 as value for this parameter which makes both objectives quite well achieved. Furthermore, the algorithm calculates the $LDL^T$ decomposition of the correlation matrix as byproduct which was used for a fast and accurate evaluation of the cost function as described in Subsection \ref{subsec: computational details}.

 \section{Results for the Application Example}
\label{sec: results}

We applied the methods for parameter estimation, uncertainty quantification and experimental design introduced in Section \ref{sec: methods} to the model for phosphate and dissolved organic phosphorus concentrations introduced in Section \ref{sec: model}. The results are presented in the following.

\subsection{Model Parameter Estimation}

We used the generalized least squares estimator, as described in Subsection \ref{subsec: parameter estimation}, to estimate the model parameters based on the measurement data described in Subsection \ref{subsec: data}. For this, the objective function was evaluated over 30.000 times with different model parameters.

Different model parameters and their cost function values are presented in Table \ref{tab: model parameters: optimal values}. The first row contains the initial guess of the model parameters presented in Subsection \ref{subsec: model parameters}. The last three rows contain the model parameters which minimize the cost function of the generalized least squares estimator (GLS), the weighted least squares estimator (WLS) and the ordinary least squares estimator (OLS), respectively. The cost function values in the table have been divided by the number of measurements to obtain values easier to interpret.

\begin{table}[H]
\centering
    \begin{tabular}{lll|llllllll}
        GLS   & WLS   & OLS     & $\kappa_{re}$  & $\alpha$  & $f_{DOP}$ & $\kappa_{\PO}$    & $\kappa_{I}$  & $k$    & $a_{re}$  & $p$\\
        \hline
1.74  & 4.30   & 0.22                       & 0.5              & 2.0       & 0.67      & 0.50              & 30            & 0.020  & 0.86        & 2.17\\
\textbf{1.22}  & 2.70   & 0.20              & 3.6               & 11.4      & 0.83      & 0.19              & 154           & 0.010  & 1.53         & 2.17\\
1.23  & \textbf{2.69}   & 0.20              & 4.7               & 10.2      & 0.88      & 0.14              & 100           & 0.011  & 1.48         & 2.19\\
1.29  & 2.88   & \textbf{0.19}              & 5.9               & 18.2      & 0.89      & 0.14              & 200           & 0.011  & 1.26         & 2.20\\
\end{tabular}
    \par\smallskip
    \caption{Cost function values for different model parameters. (GLS: generalized least squares estimator , WLS: weighted least squares estimator, OLS: ordinary least squares estimator)}
    \label{tab: model parameters: optimal values}
\end{table}

Depending on the estimator, the optimal model parameters vary. However, all are better than the initial guess regardless of which estimator is considered.

We focus, as before, on the generalized least squares estimator and the corresponding optimal model parameters. They differ significantly in some cases compared to their initial guess: $\kappa_{re}$, $\alpha$ and $\kappa_{I}$ are significantly higher. $f_{DOP}$ and $a_{re}$ are slightly higher. $\kappa_{\PO}$ and $k$ are slightly lower. $p$ is equal to the initial guess which is reasonable since this parameter could already be estimated very well directly from the measurement data.

If the statistical assumption \eqref{eq: Y: assumed distribution}, on which the generalized least squares estimator is based is correct, the expected value of corresponding estimator divided by the number of measurements is equal to one. In our case it is approximately one point two which indicates that the assumptions might be correct or at least not too far from reality. This in turn confirms our approach.

\begin{figure}[t]
    \begin{minipage}{0.50\textwidth}
    	\begin{subfigure}{1\linewidth}
    		\includegraphics[width=1.0\linewidth]{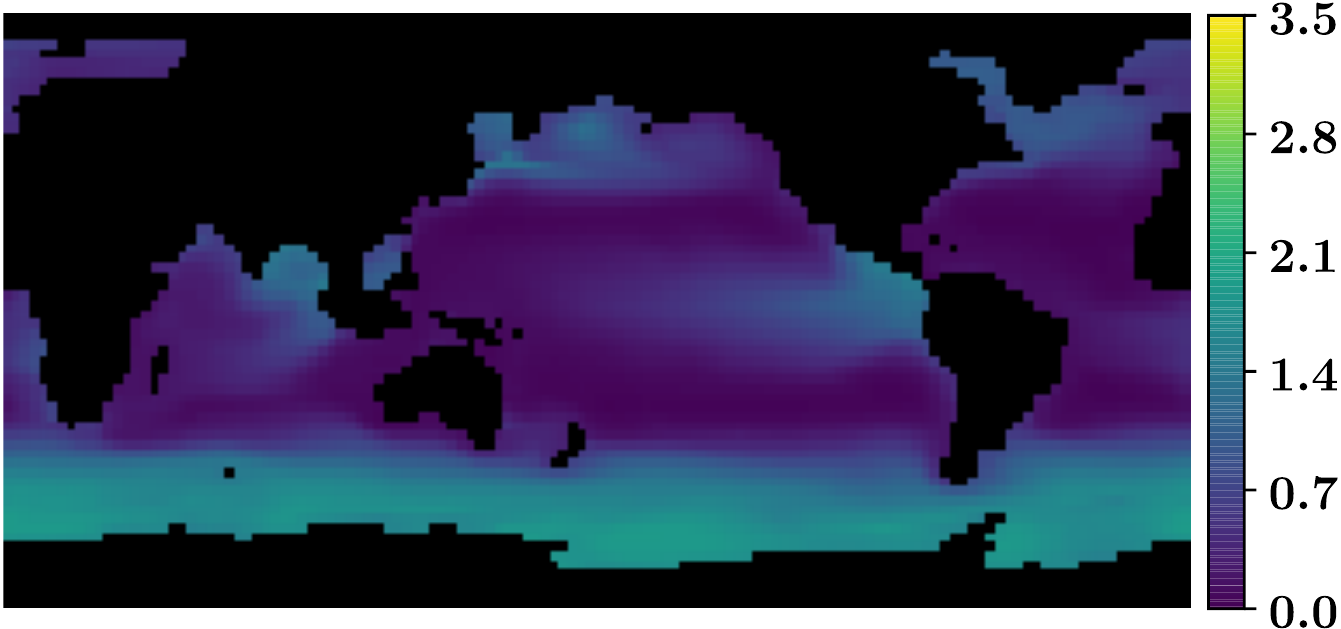}
       		\caption{water surface: averaged over time and 0 to 25 ${\rm m}$ depth} 
    		\label{fig:model_output:po4:surface}
    	\end{subfigure}
        \par\smallskip
    	\begin{subfigure}{0.49\linewidth}
    		\includegraphics[width=1.0\linewidth, height=0.8\linewidth]{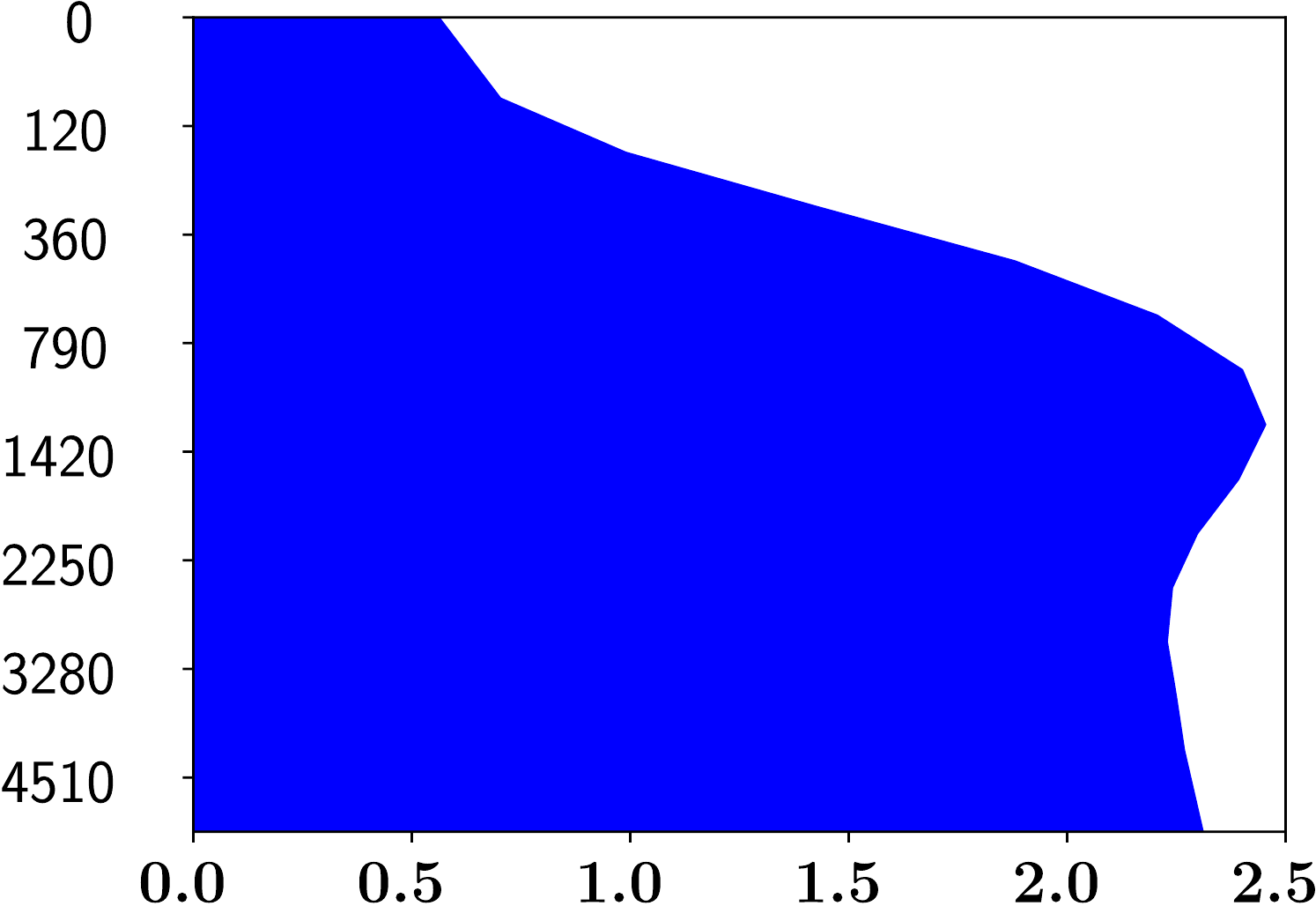}
    		\caption{averaged over all but depth} 
    		\label{fig:model_output:po4:depth}
    	\end{subfigure}
       	\hfill
       	\begin{subfigure}{0.49\linewidth}
       		\includegraphics[width=1.0\linewidth, height=0.8\linewidth]{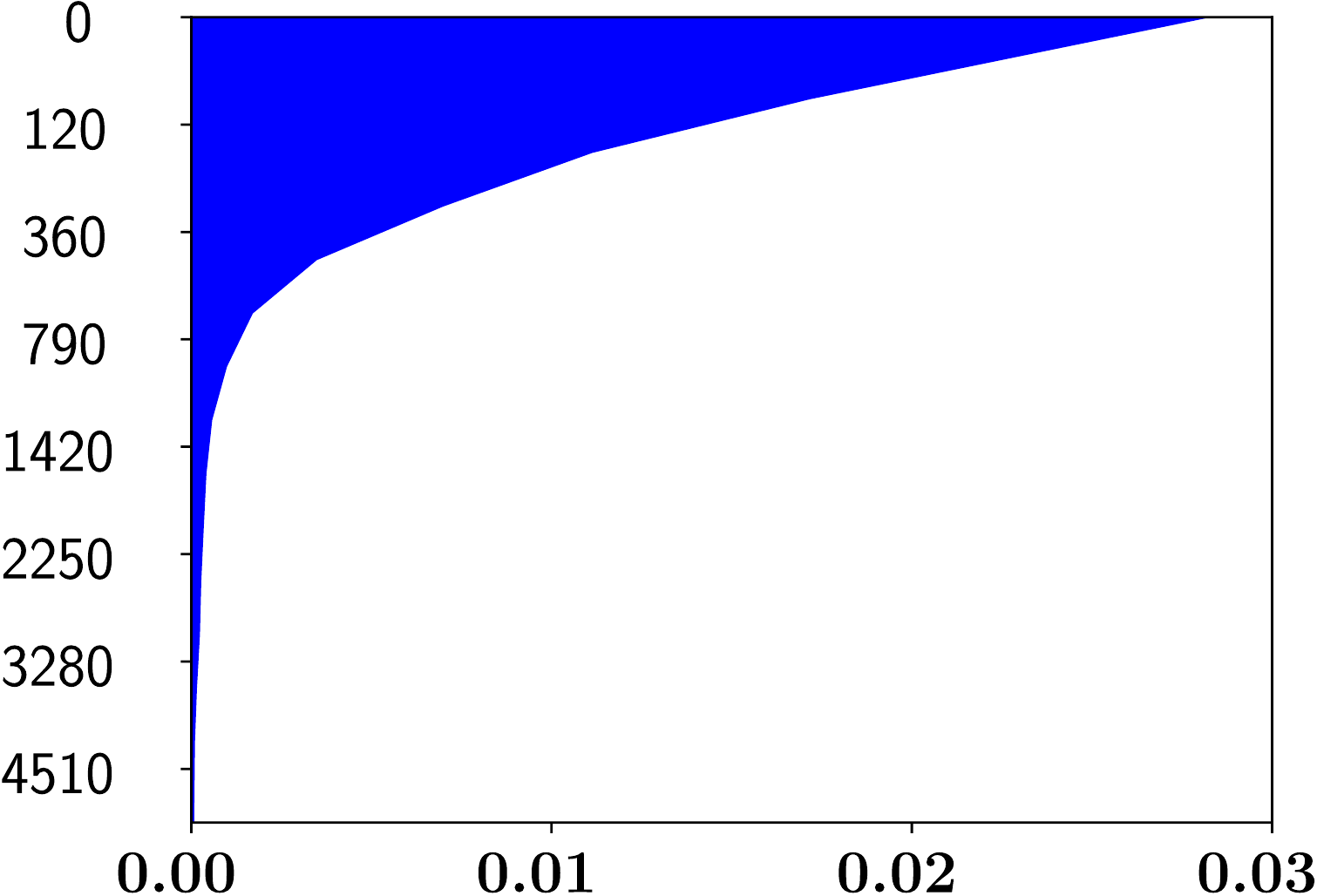}
       		\caption{average monthly change}
       		\label{fig:model_output:po4:time_diff}
       	\end{subfigure}
    \end{minipage}
    \hfill
    \begin{minipage}{0.49\textwidth}
       	\begin{subfigure}{1\linewidth}
       		\includegraphics[width=1.0\linewidth]{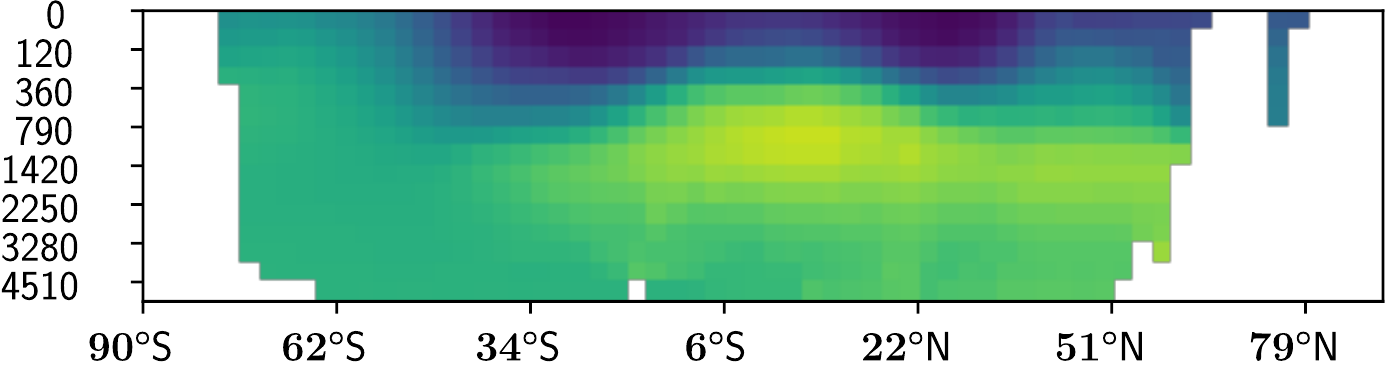}
       		\caption{Pacific Ocean: averaged over time and between 125$\degree$E and 70$\degree$W}
       		\label{fig:model_output:po4:pacific}
       	\end{subfigure}
        \par\smallskip
       	\begin{subfigure}{1\linewidth}
       		\includegraphics[width=1.0\linewidth]{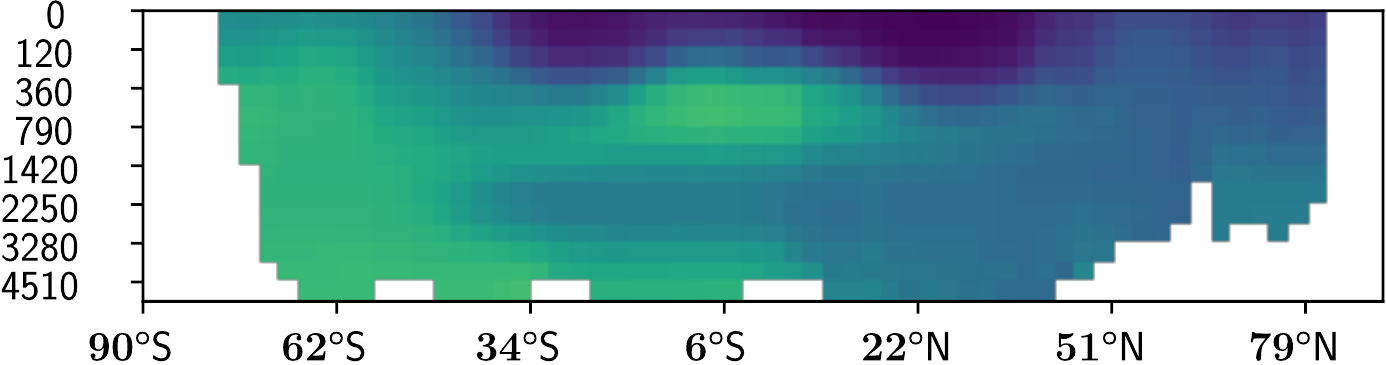}
       		\caption{Atlantic Ocean: averaged over time and between 70$\degree$W and 20$\degree$E} 
       		\label{fig:model_output:po4:atlantic}
       	\end{subfigure}
        \par\smallskip
       	\begin{subfigure}{1\linewidth}
       		\includegraphics[width=1.0\linewidth]{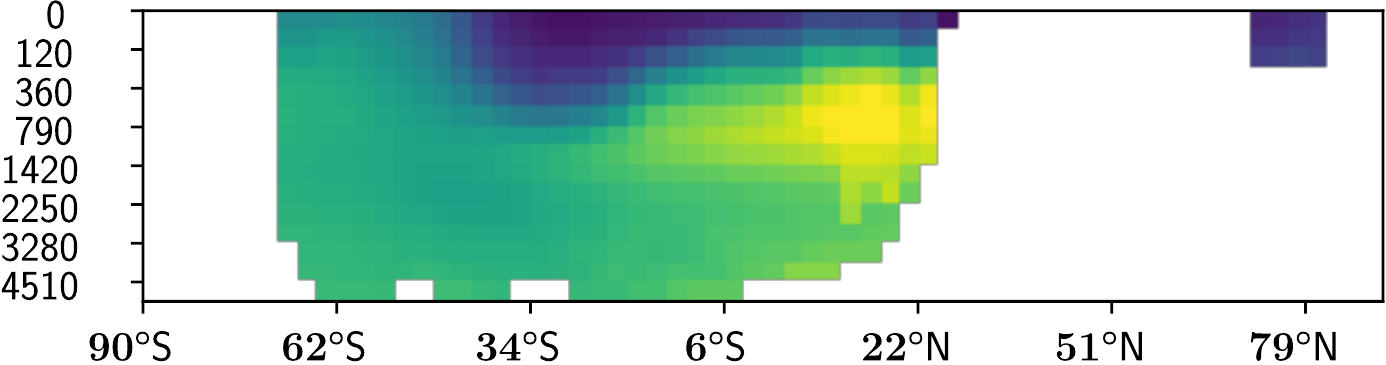}
       		\caption{Indian Ocean: averaged over time and between 20$\degree$W and 125$\degree$E}
       		\label{fig:model_output:po4:indian}
       	\end{subfigure}
    \end{minipage}
   	\caption{Model output for phosphate (in ${\rm mmol\,m}^{-3}$) with model parameters estimated by GLS.}
	\label{fig:model_output:po4} 
\end{figure}

\begin{figure}[t]
    \begin{minipage}{0.50\textwidth}
    	\begin{subfigure}{1\linewidth}
    		\includegraphics[width=1.0\linewidth]{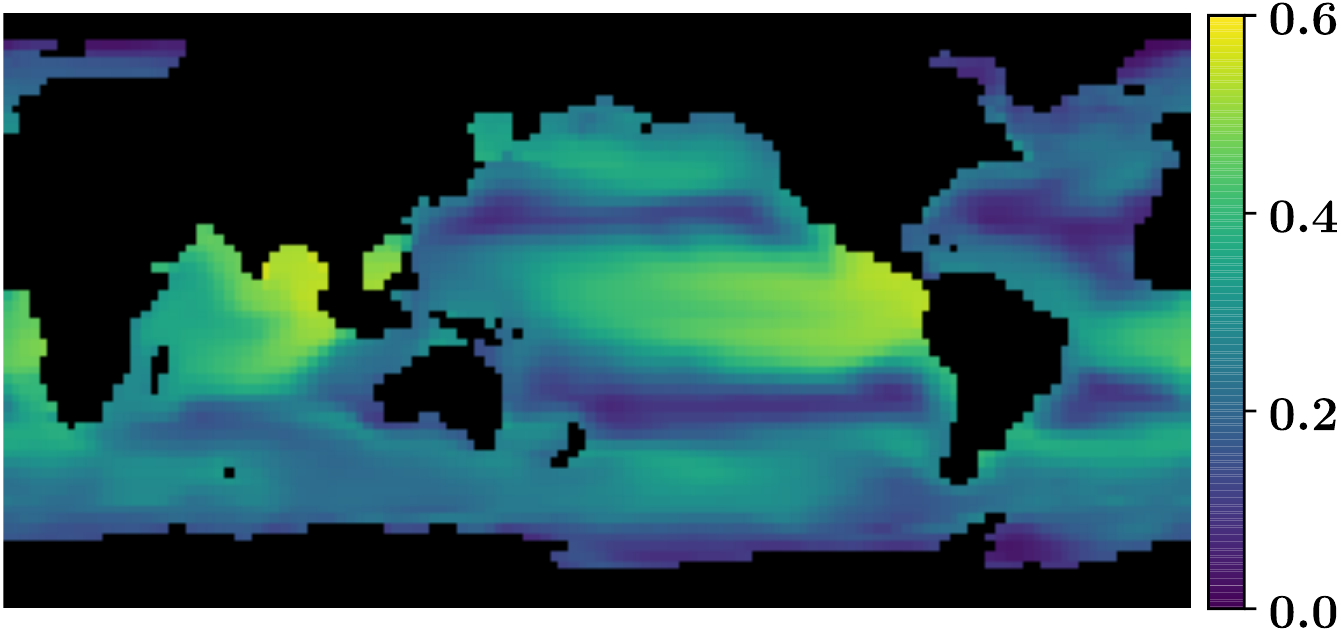}
       		\caption{water surface: averaged over time and 0 to 25 ${\rm m}$ depth} 
    		\label{fig:model_output:dop:surface}
    	\end{subfigure}
        \par\smallskip
    	\begin{subfigure}{0.49\linewidth}
    		\includegraphics[width=1.0\linewidth, height=0.8\linewidth]{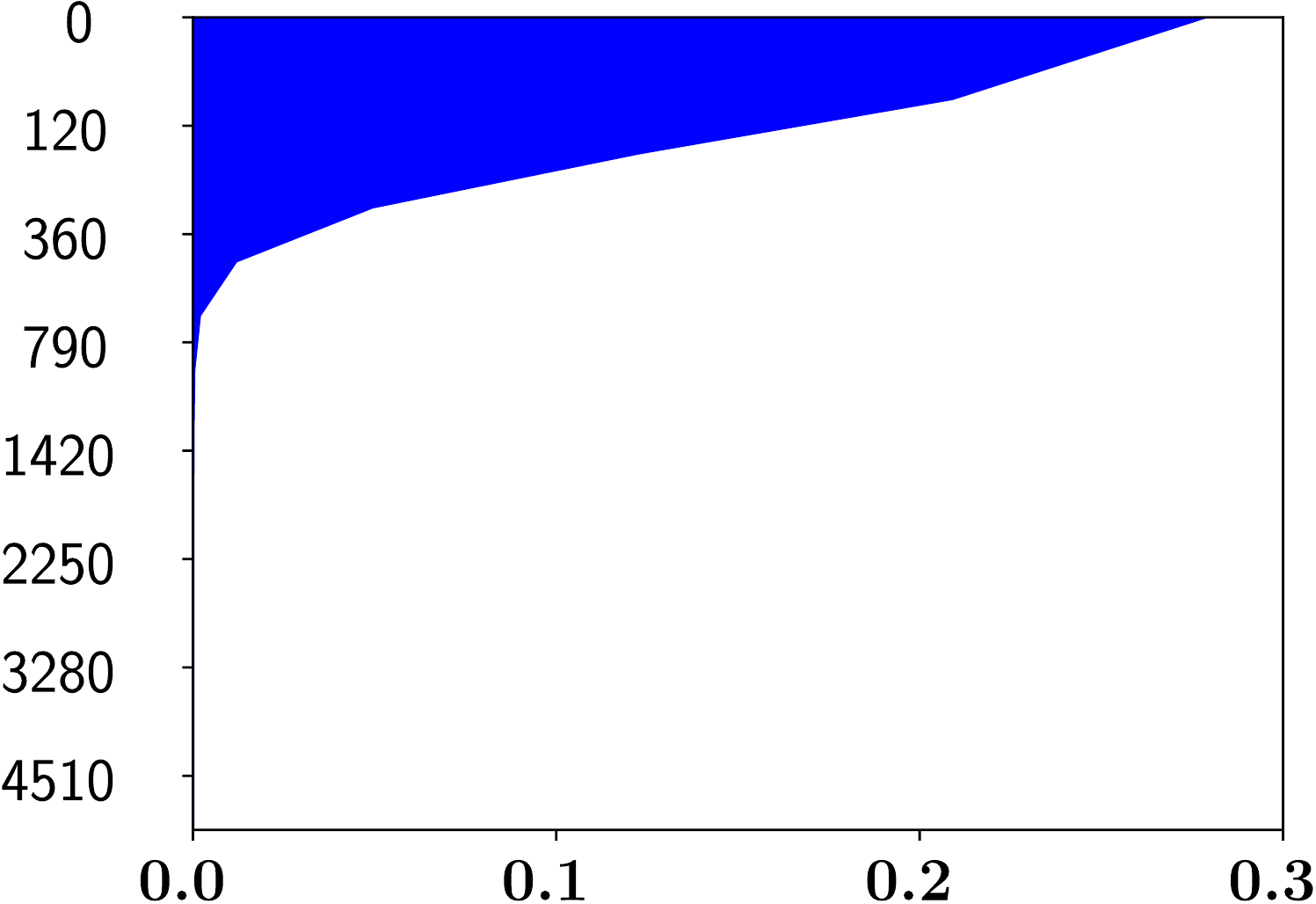}
    		\caption{averaged over all but depth} 
    		\label{fig:model_output:dop:depth}
    	\end{subfigure}
       	\hfill
       	\begin{subfigure}{0.49\linewidth}
       		\includegraphics[width=1.0\linewidth, height=0.8\linewidth]{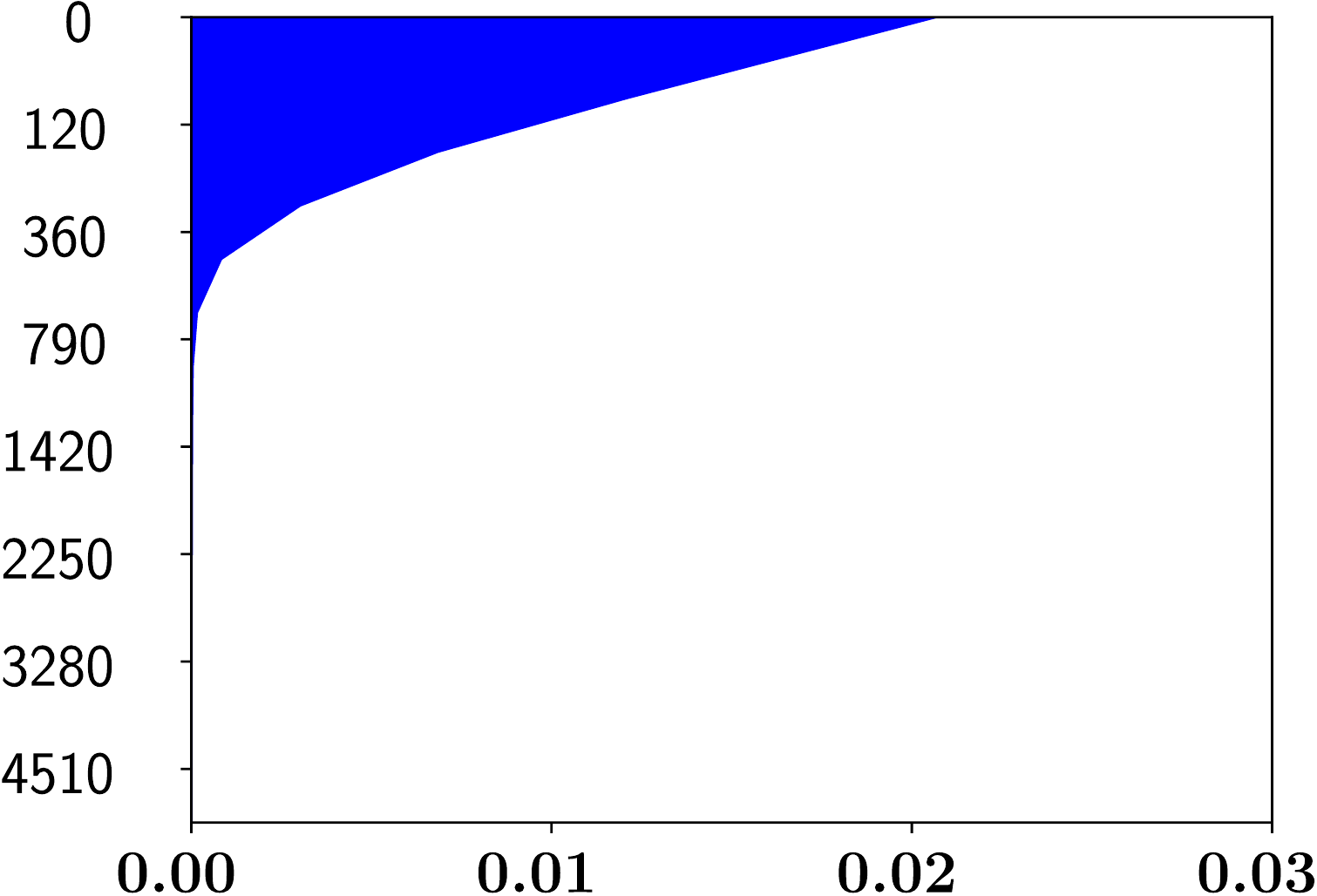}
       		\caption{average monthly change}
       		\label{fig:model_output:dop:time_diff}
       	\end{subfigure}
    \end{minipage}
    \hfill
    \begin{minipage}{0.49\textwidth}
       	\begin{subfigure}{1\linewidth}
       		\includegraphics[width=1.0\linewidth]{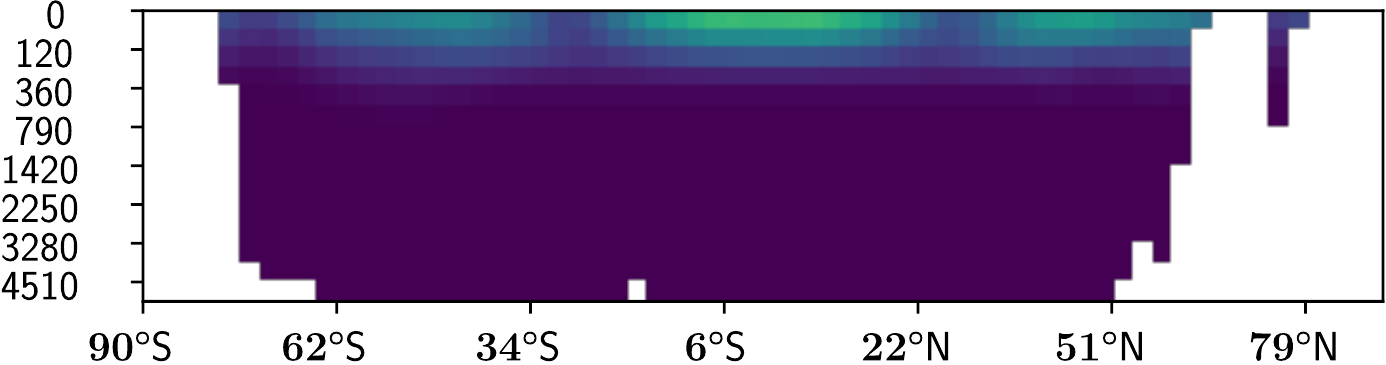}
       		\caption{Pacific Ocean: averaged over time and between 125$\degree$E and 70$\degree$W} 
       		\label{fig:model_output:dop:pacific}
       	\end{subfigure}
        \par\smallskip
       	\begin{subfigure}{1\linewidth}
       		\includegraphics[width=1.0\linewidth]{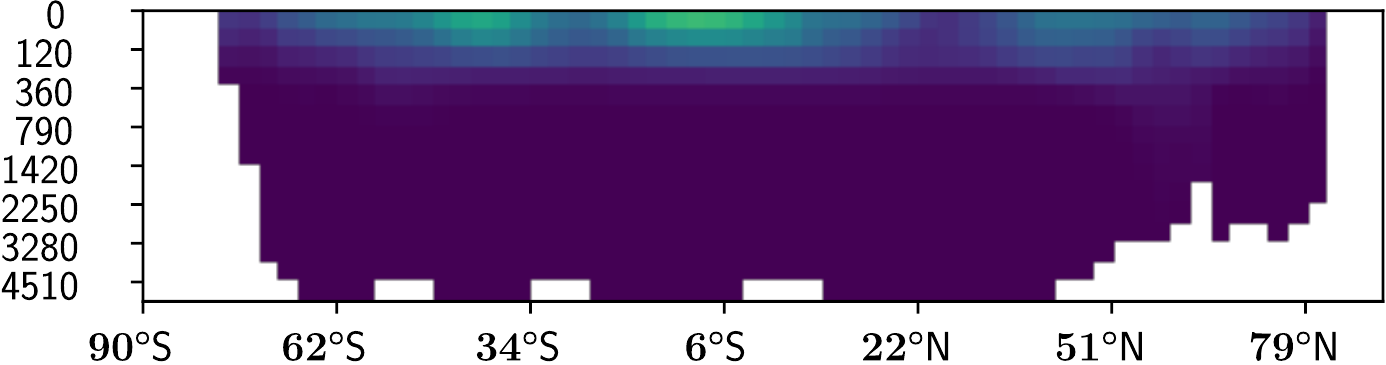}
       		\caption{Atlantic Ocean: averaged over time and between 70$\degree$W and 20$\degree$E} 
       		\label{fig:model_output:dop:atlantic}
       	\end{subfigure}
        \par\smallskip
       	\begin{subfigure}{1\linewidth}
       		\includegraphics[width=1.0\linewidth]{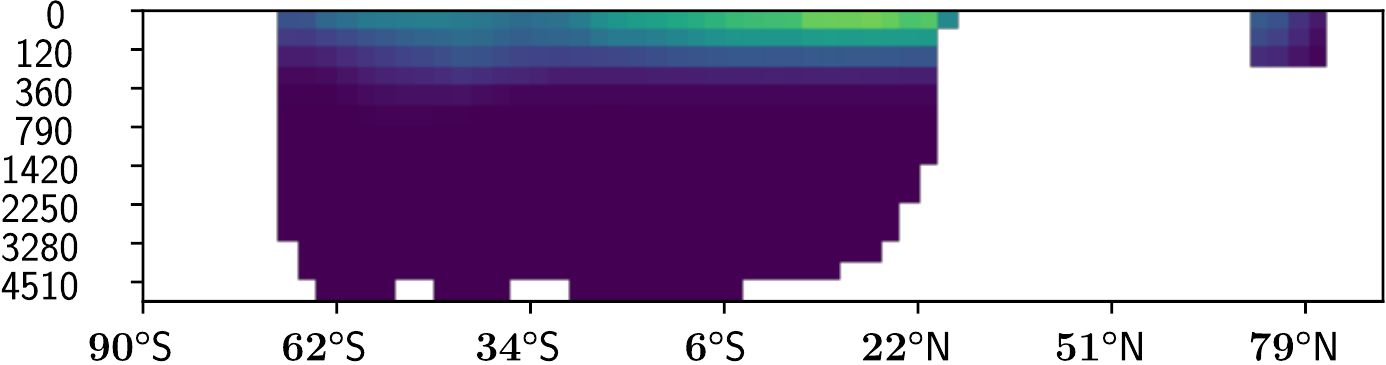}
       		\caption{Indian Ocean: averaged over time and between 20$\degree$W and 125$\degree$E}
       		\label{fig:model_output:dop:indian}
       	\end{subfigure}
    \end{minipage}
   	\caption{Model output for dissolved organic phosphorus (in ${\rm mmol\,m}^{-3}$) with model parameters estimated by GLS.}
	\label{fig:model_output:dop} 
\end{figure}

The model output with the optimal model parameters regarding the generalized least squares estimator (second row in Table \ref{tab: model parameters: optimal values}) is summarized in Figure \ref{fig:model_output:po4} and \ref{fig:model_output:dop}. The time averaged output at the water surface is plotted in Figure \ref{fig:model_output:po4:surface} and \ref{fig:model_output:dop:surface}. The average model output depending on the depth is shown in Figure \ref{fig:model_output:po4:depth} and \ref{fig:model_output:dop:depth}. The average absolute change after one month is plotted in Figure \ref{fig:model_output:po4:time_diff} and \ref{fig:model_output:dop:time_diff}. Figure \ref{fig:model_output:po4:pacific} and \ref{fig:model_output:dop:pacific}, \ref{fig:model_output:po4:atlantic} and \ref{fig:model_output:dop:atlantic} as well as \ref{fig:model_output:po4:indian} and \ref{fig:model_output:dop:indian} show the model output in the Pacific Ocean, the Atlantic Ocean and the Indian Ocean, respectively, depending on depth and latitude and averaged over time and between the corresponding longitudes.

\vspace{-0.2em}

The average phosphate concentration at the surface is roughly 0.6 ${\rm mmol\,m}^{-3}$. It increases with growing depth. Deeper than 700 meters the average is approximate constant 2.3 ${\rm mmol\,m}^{-3}$.

The temporal variability decreases with growing depth. The average monthly change of the concentrations is around 0.03 ${\rm mmol\,m}^{-3}$ at the surface. There are almost no changes over time deeper than 700 m.

At the water surface, the highest concentrations are at the Southern Ocean with around 2.2 ${\rm mmol\,m}^{-3}$ and at the north-eastern part of the Indian Ocean, the northern and middle-east part of the Pacific Ocean as well as the northern part of the Atlantic Ocean ranging from 1 to 2 ${\rm mmol\,m}^{-3}$.

The phosphate concentration is highest in each of the Pacific Ocean, the Atlantic Ocean and the Indian Ocean around the equator at a depth between 500 and 1500 meters. The lowest concentrations in each of these oceans is around the equator near the water surface.

The average dissolved organic phosphorus concentration is almost 0.3 ${\rm mmol\,m}^{-3}$ at the water surface and decreases quickly with growing depth. It is close to zero below 500 m.

The temporal variability decreases rapidly as well with growing depth. The average monthly change of the concentrations is around 0.02 ${\rm mmol\,m}^{-3}$ at the surface and there are almost no changes over time below 500 m.

The highest dissolved organic phosphorus concentrations of almost 0.6 ${\rm mmol\,m}^{-3}$ are at the surface around the equator. Other high values with around 0.4 ${\rm mmol\,m}^{-3}$ are in areas around 45$\degree$S and 45$\degree$N.

The previously described behavior applies to the Pacific Ocean, the Atlantic Ocean as well as the Indian Ocean.

\subsection{Uncertainty in Parameter Estimation}

The uncertainty in the parameter estimation has been quantified as described in Subsection \ref{subsec: parameter uncertainty}. For this, we have approximated the covariance matrix of the parameter estimator and confidence intervals with a confidence level of 99\% using Equation \eqref{eq: P: covariance matrix: Jacobian and Hessian} and \eqref{eq: confidence intervals for p with estimated sigma}.

\begin{figure}[H]
\begin{minipage}[t]{0.49\textwidth}
    \includegraphics[width=\textwidth,height=0.8\textwidth]{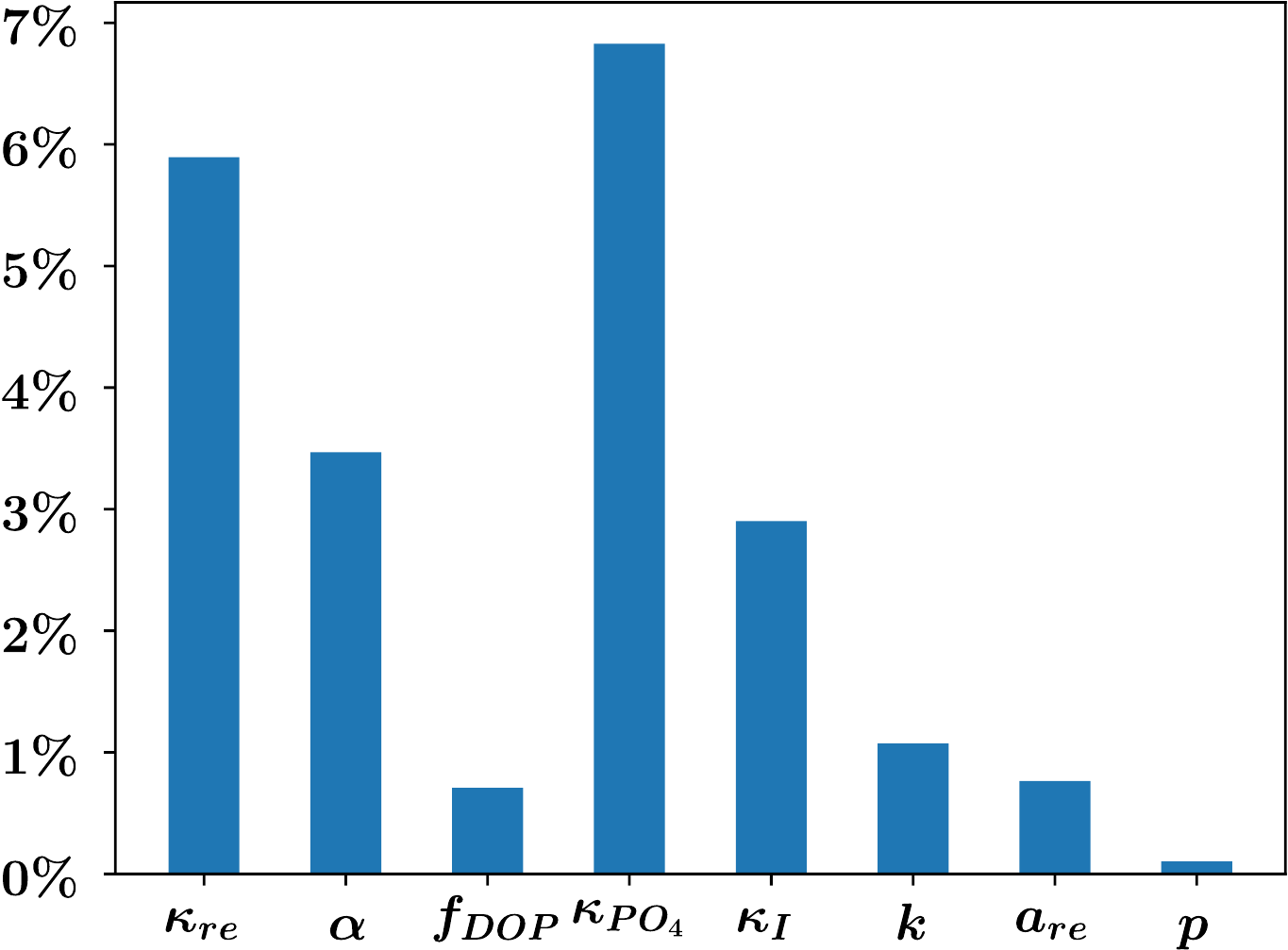}
   \caption{Confidence intervals length relative to estimated model parameters with 99 \% confidence level.} 
    \label{fig:uncertainty_parameters:confidence}
\end{minipage}
\begin{minipage}[t]{0.49\textwidth}
    \includegraphics[width=\textwidth]{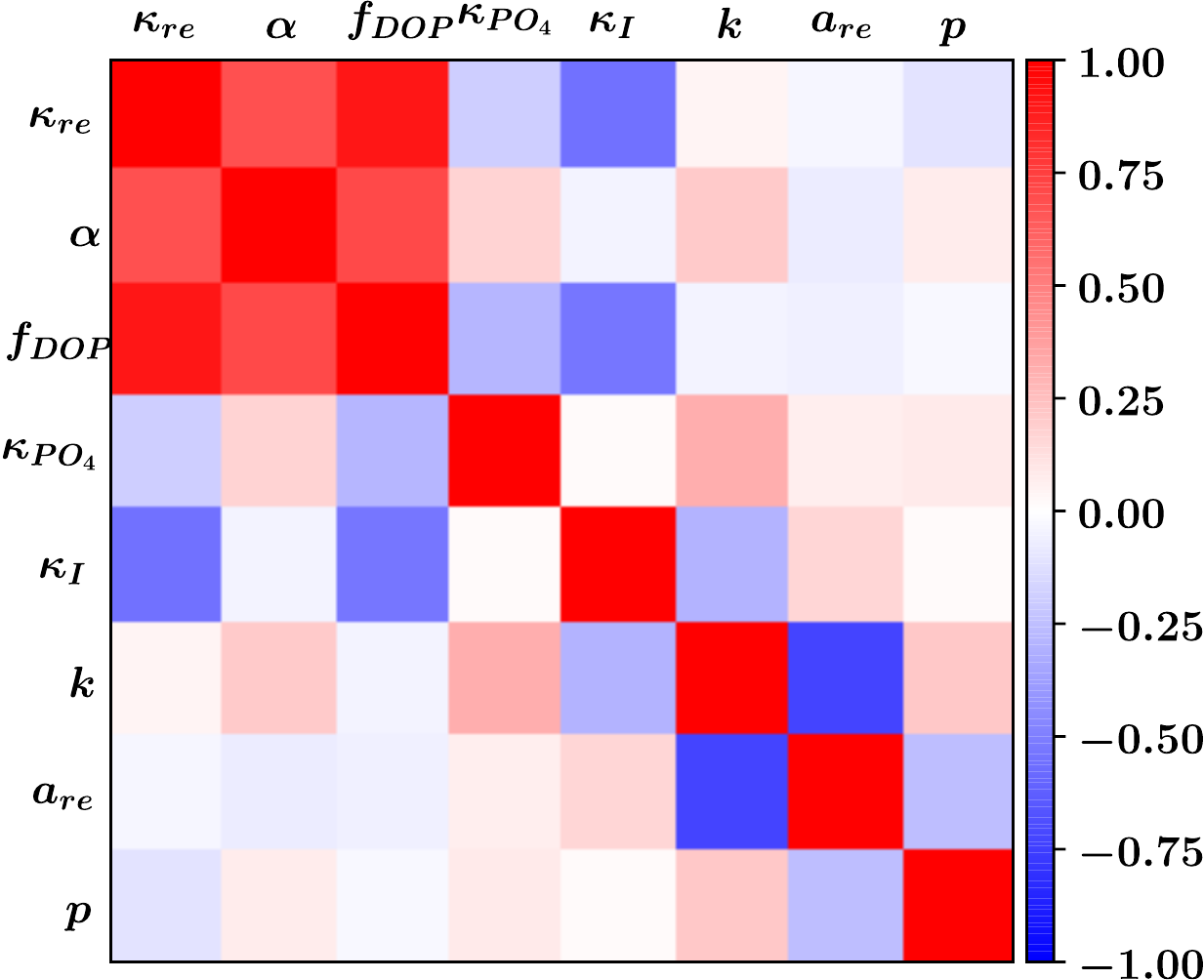}
    \caption{Correlation matrix of the model parameter estimator (generalized least squares estimator).} 
    \label{fig:uncertainty_parameters:correlations}
\end{minipage}
\end{figure}

For each model parameter, the length of its confidence interval relative to its estimated value is plotted in Figure \ref{fig:uncertainty_parameters:confidence}. The estimates with the greatest uncertainty are those for $\kappa_{re}$ and $\kappa_{\PO}$ with six to seven percent. These are followed by $\alpha$ and $\kappa_{I}$ with around three percent. A lower uncertainty of about one percent is associated with $f_{DOP}$, $k$ and $a_{re}$. The slightest uncertainty of one per mill is associated with $p$.

These values are consistent with our experience with the model. Its output is sensitive to changes in the parameters $f_{DOP}$, $k$ and $a_{re}$ and very sensitive to changes in $p$. Hence, it is reasonable that these parameter could be estimated quite accurately.

The correlation matrix of the parameters estimator is plotted in Figure \ref{fig:uncertainty_parameters:correlations}. Here strong positive correlations between $\kappa_{re}$, $\alpha$ and $f_{DOP}$ are conspicuous. They imply that if the true value of one of these model parameters is higher or lower than its estimate, it is very likely that the same applies to the other two parameters. Especially $\kappa_{re}$ and $f_{DOP}$ have a correlation close to one.

A strong negative correlation close to minus one is between $k$ and $a_{re}$. This means that if the true value of one of these parameters is greater than its estimate, then it is very likely that the other is smaller and vice versa.

We also compared the different approaches to approximated the covariance matrix of the parameter estimator described in Equation \eqref{eq: P: covariance matrix: Jacobian}, \eqref{eq: P: covariance matrix: Hessian} and \eqref{eq: P: covariance matrix: Jacobian and Hessian}. All three approximations provide similar results. They differ in each component usually at most by a factor between one half and two. The similarity of the three approximations supports the statistical assumption \eqref{eq: Y: assumed distribution}.

\subsection{Uncertainty in Model Output}

\begin{figure}[t]
    \begin{minipage}{0.50\textwidth}
       	\begin{subfigure}{1\linewidth}
       		\includegraphics[width=1.0\linewidth]{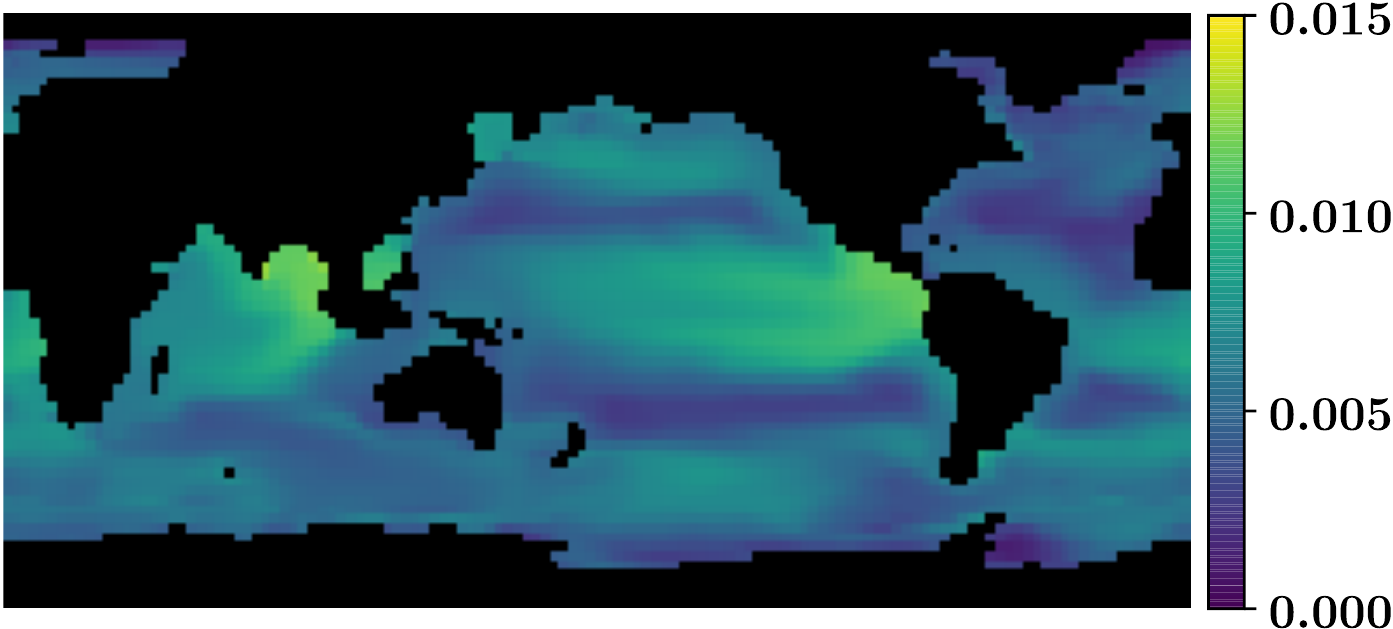}
          		\caption{water surface: averaged over time and 0 to 25 ${\rm m}$ depth} 
       		\label{fig:model_confidence:po4:surface}
       	\end{subfigure}
        \par\smallskip
    	\begin{subfigure}{0.49\linewidth}
    		\includegraphics[width=1.0\linewidth, height=0.8\linewidth]{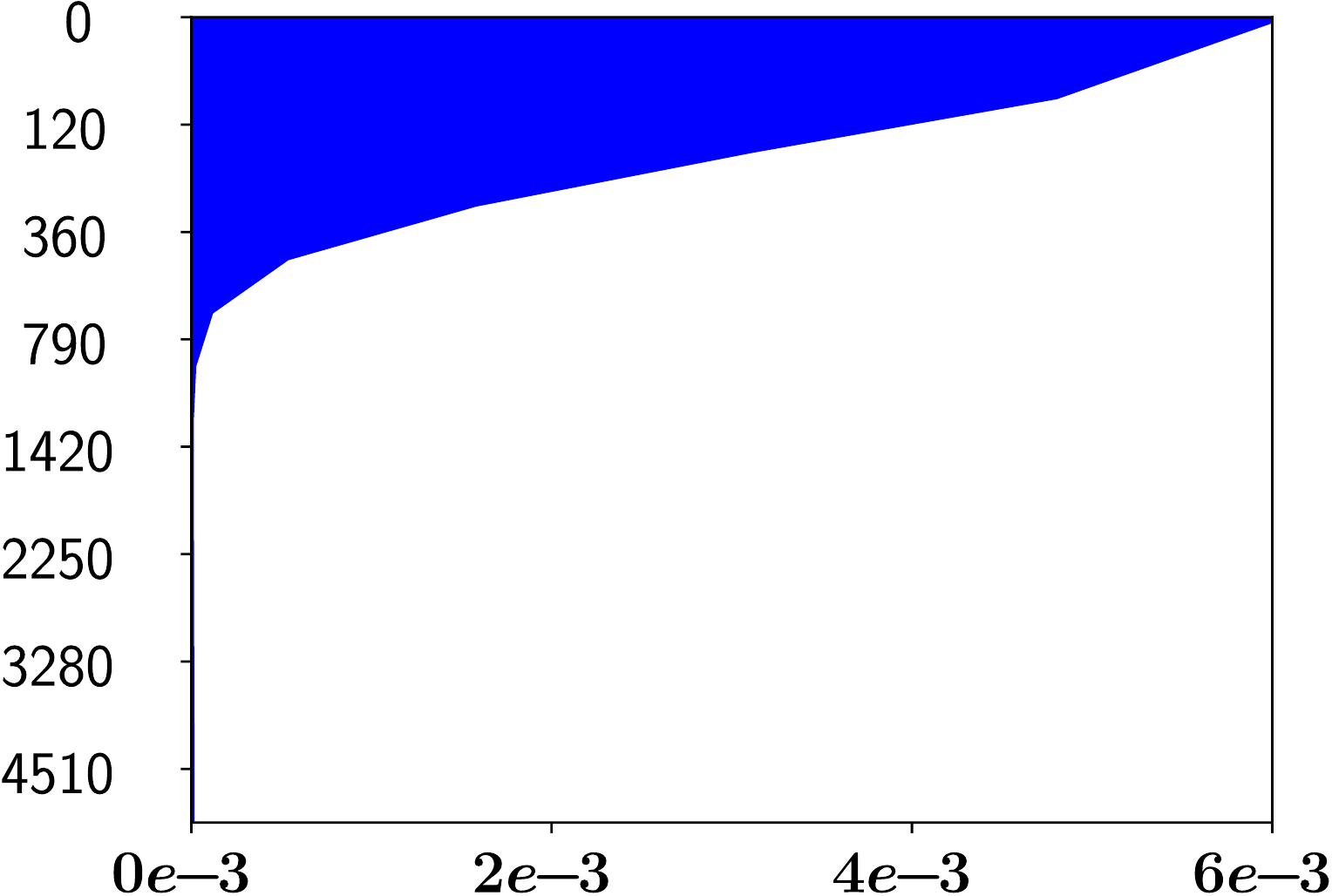}
    		\caption{averaged over all but depth} 
    		\label{fig:model_confidence:po4:depth}
    	\end{subfigure}
       	\hfill
       	\begin{subfigure}{0.49\linewidth}
       		\includegraphics[width=1.0\linewidth, height=0.8\linewidth]{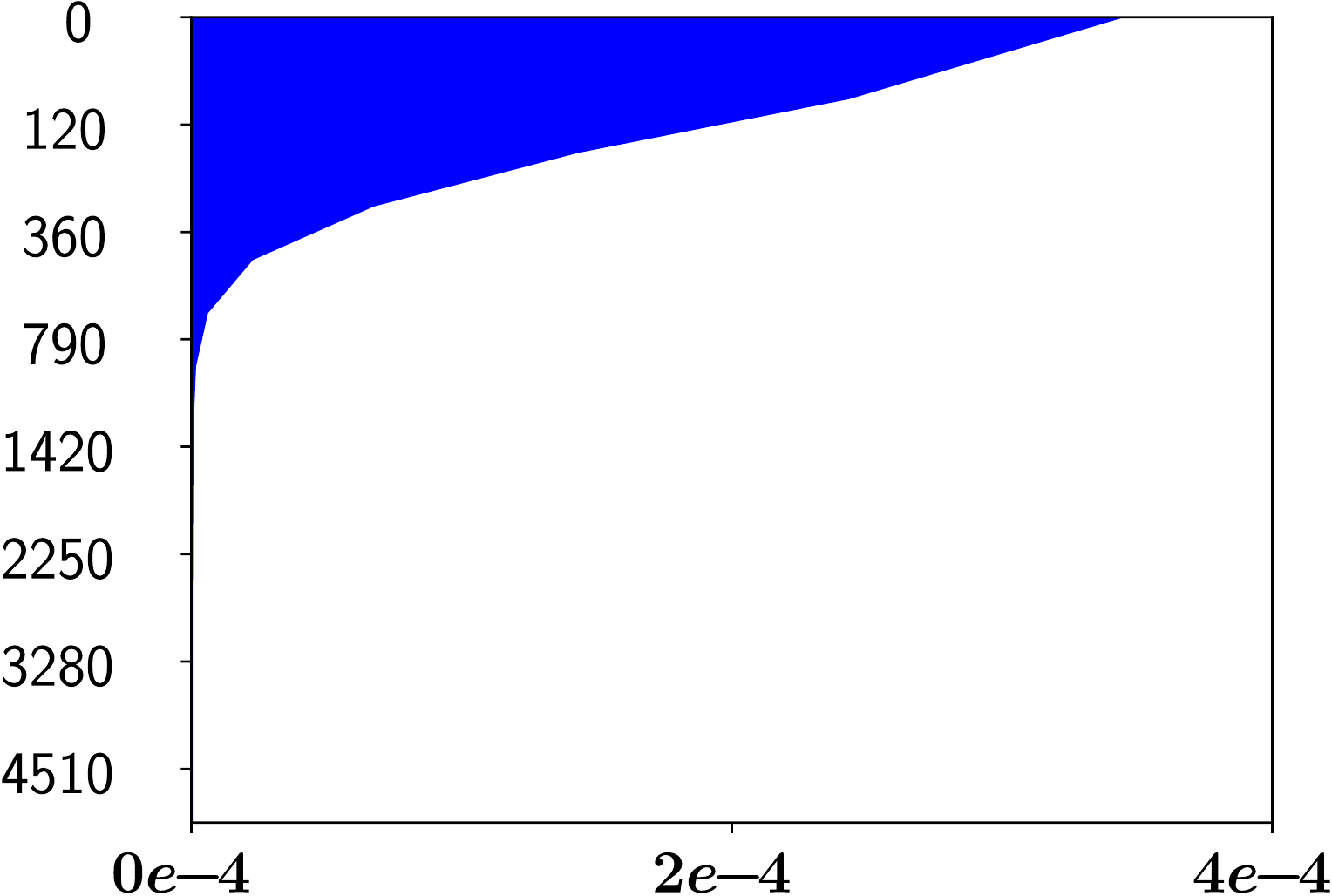}
       		\caption{average monthly change}
       		\label{fig:model_confidence:po4:time_diff}
       	\end{subfigure}
    \end{minipage}
    \hfill
    \begin{minipage}{0.49\textwidth}
       	\begin{subfigure}{1\linewidth}
       		\includegraphics[width=1.0\linewidth]{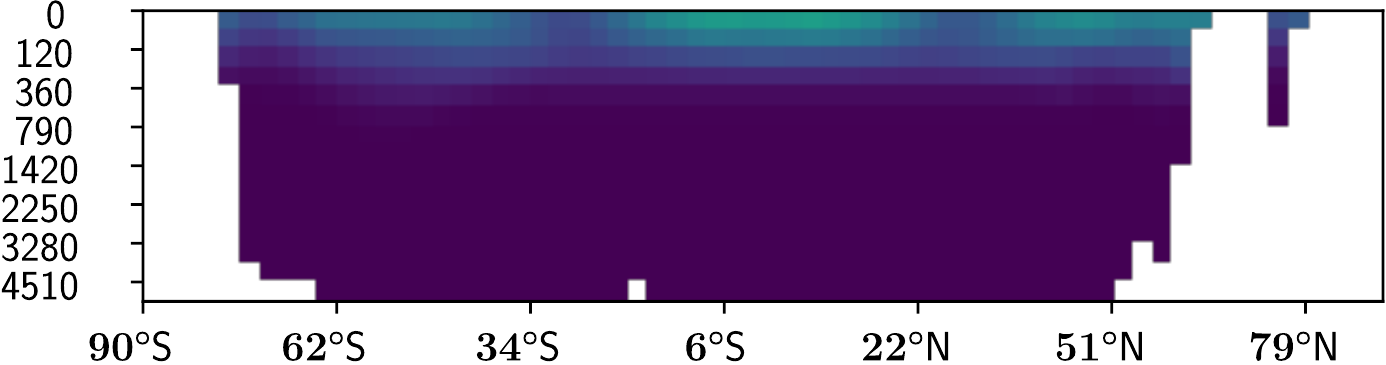}
       		\caption{Pacific Ocean: averaged over time and between 125$\degree$E and 70$\degree$W}
       		\label{fig:model_confidence:po4:pacific}
       	\end{subfigure}
        \par\smallskip
       	\begin{subfigure}{1\linewidth}
       		\includegraphics[width=1.0\linewidth]{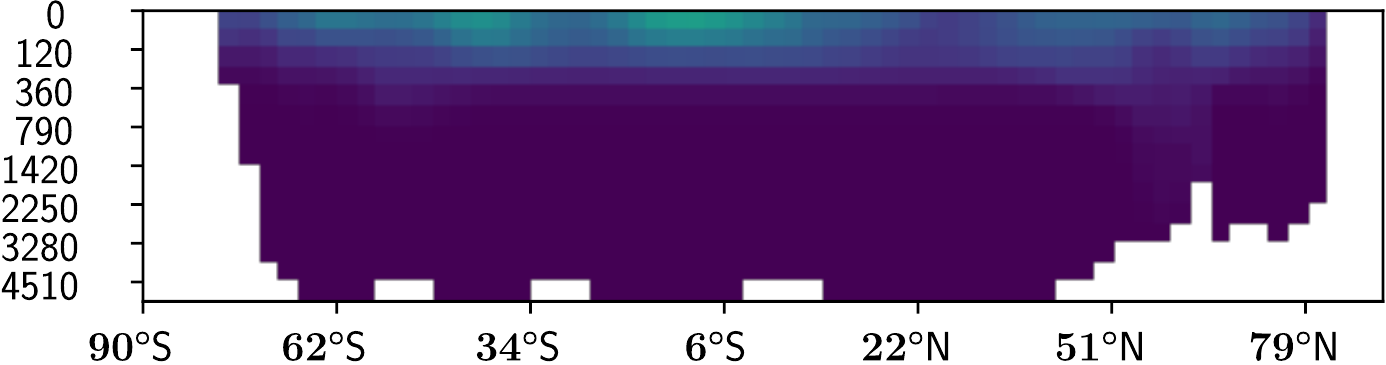}
       		\caption{Atlantic Ocean: averaged over time and between 70$\degree$W and 20$\degree$E} 
       		\label{fig:model_confidence:po4:atlantic}
       	\end{subfigure}
        \par\smallskip
       	\begin{subfigure}{1\linewidth}
       		\includegraphics[width=1.0\linewidth]{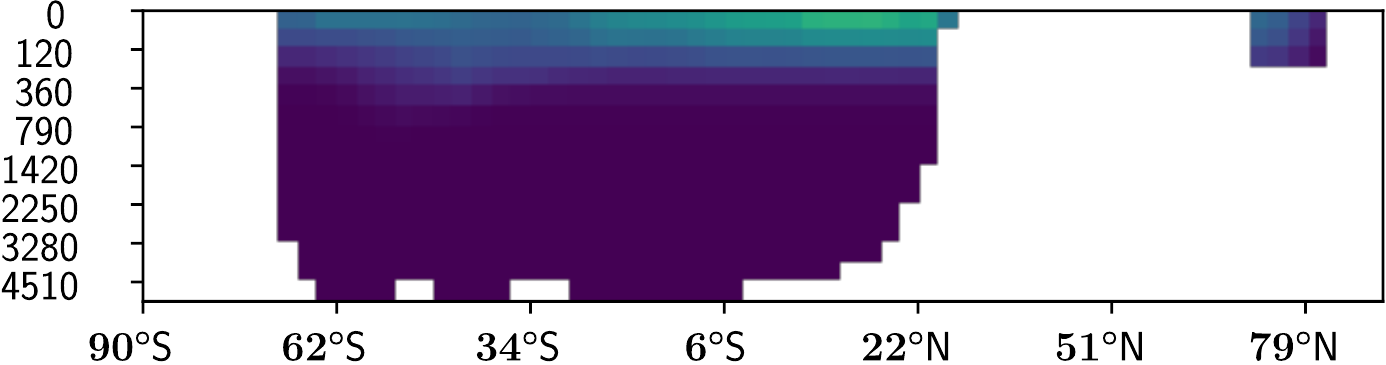}
       		\caption{Indian Ocean: averaged over time and between 20$\degree$W and 125$\degree$E}
       		\label{fig:model_confidence:po4:indian}
       	\end{subfigure}
    \end{minipage}
    \caption{Confidence intervals length for phosphate model output (in ${\rm mmol\,m}^{-3}$) with 99 \% confidence level.}
	\label{fig:model_confidence:po4} 
\end{figure}

\begin{figure}[t]
    \begin{minipage}{0.50\textwidth}
    	\begin{subfigure}{1\linewidth}
    		\includegraphics[width=1.0\linewidth]{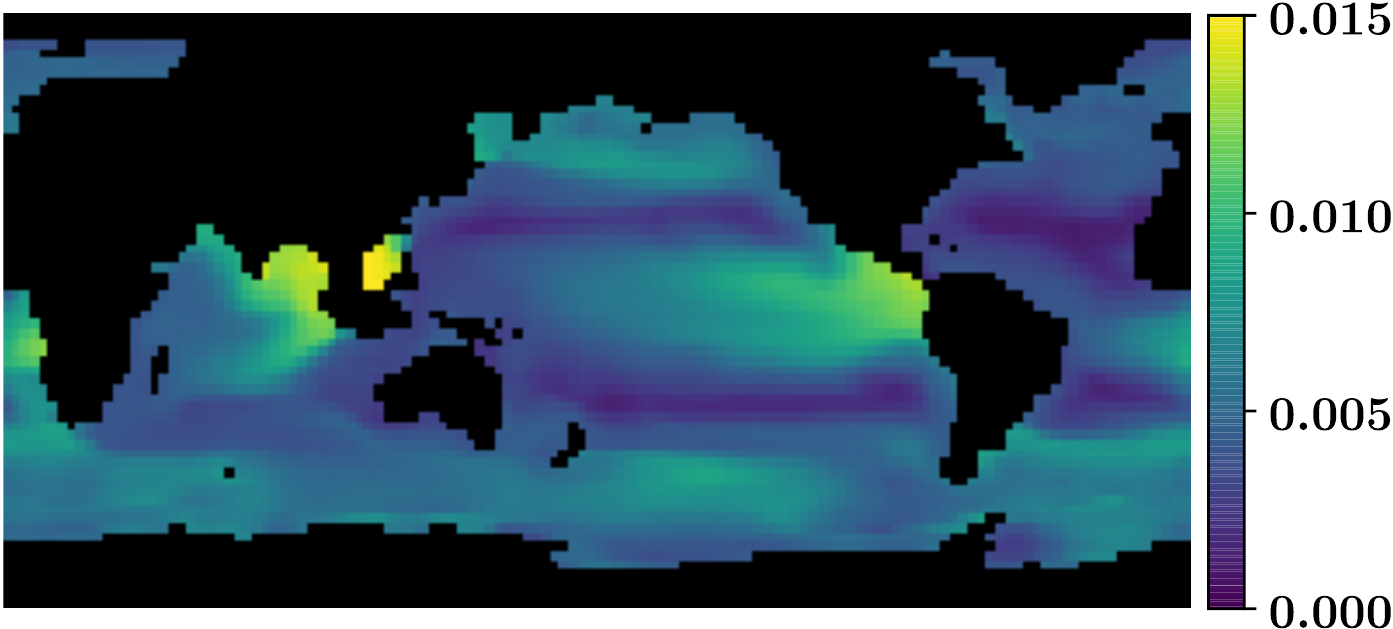}
       		\caption{water surface: averaged over time and 0 to 25 ${\rm m}$ depth} 
    		\label{fig:model_confidence:dop:surface}
    	\end{subfigure}
        \par\smallskip
    	\begin{subfigure}{0.49\linewidth}
    		\includegraphics[width=1.0\linewidth, height=0.8\linewidth]{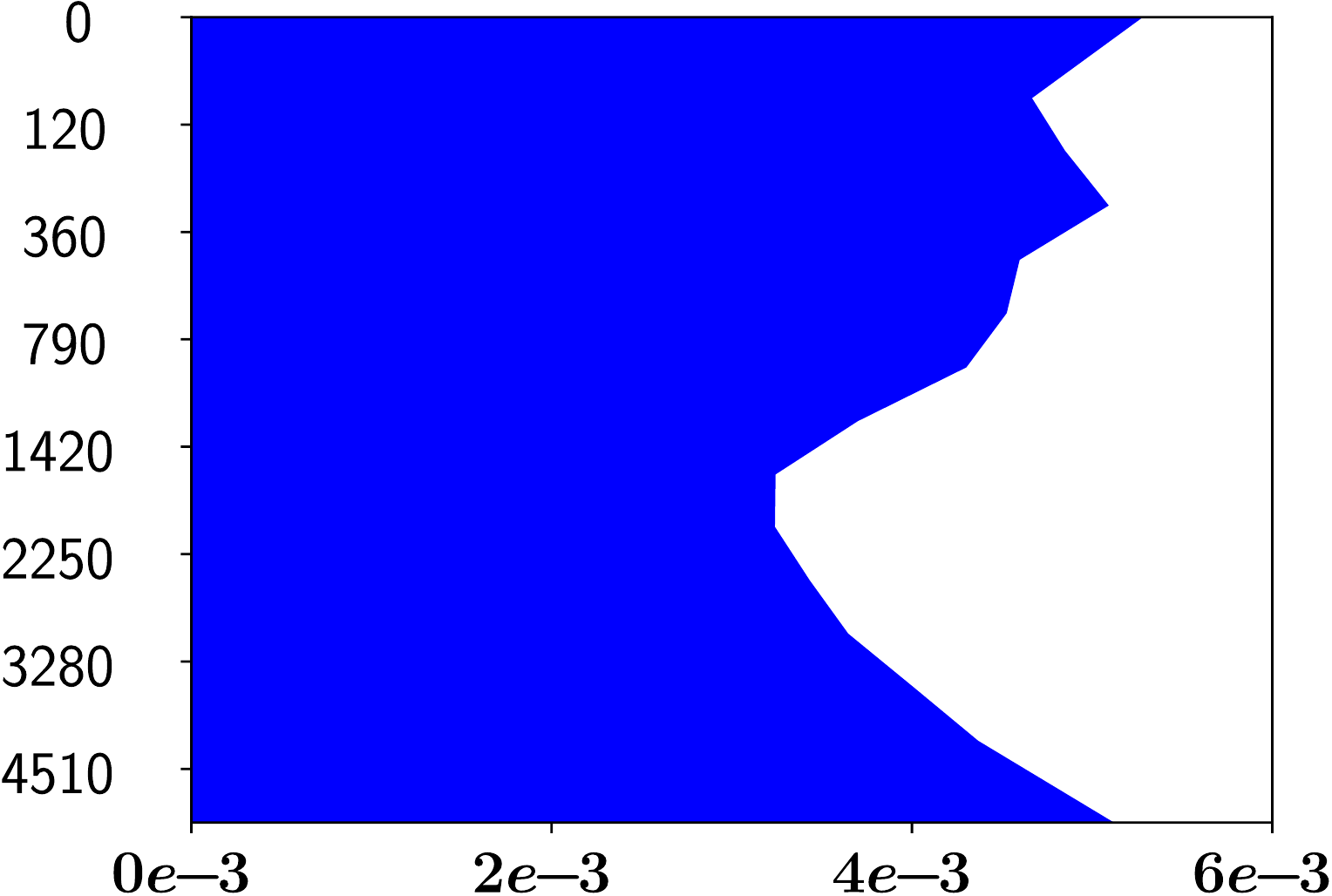}
    		\caption{averaged over all but depth} 
    		\label{fig:model_confidence:dop:depth}
    	\end{subfigure}
       	\hfill
       	\begin{subfigure}{0.49\linewidth}
       		\includegraphics[width=1.0\linewidth, height=0.8\linewidth]{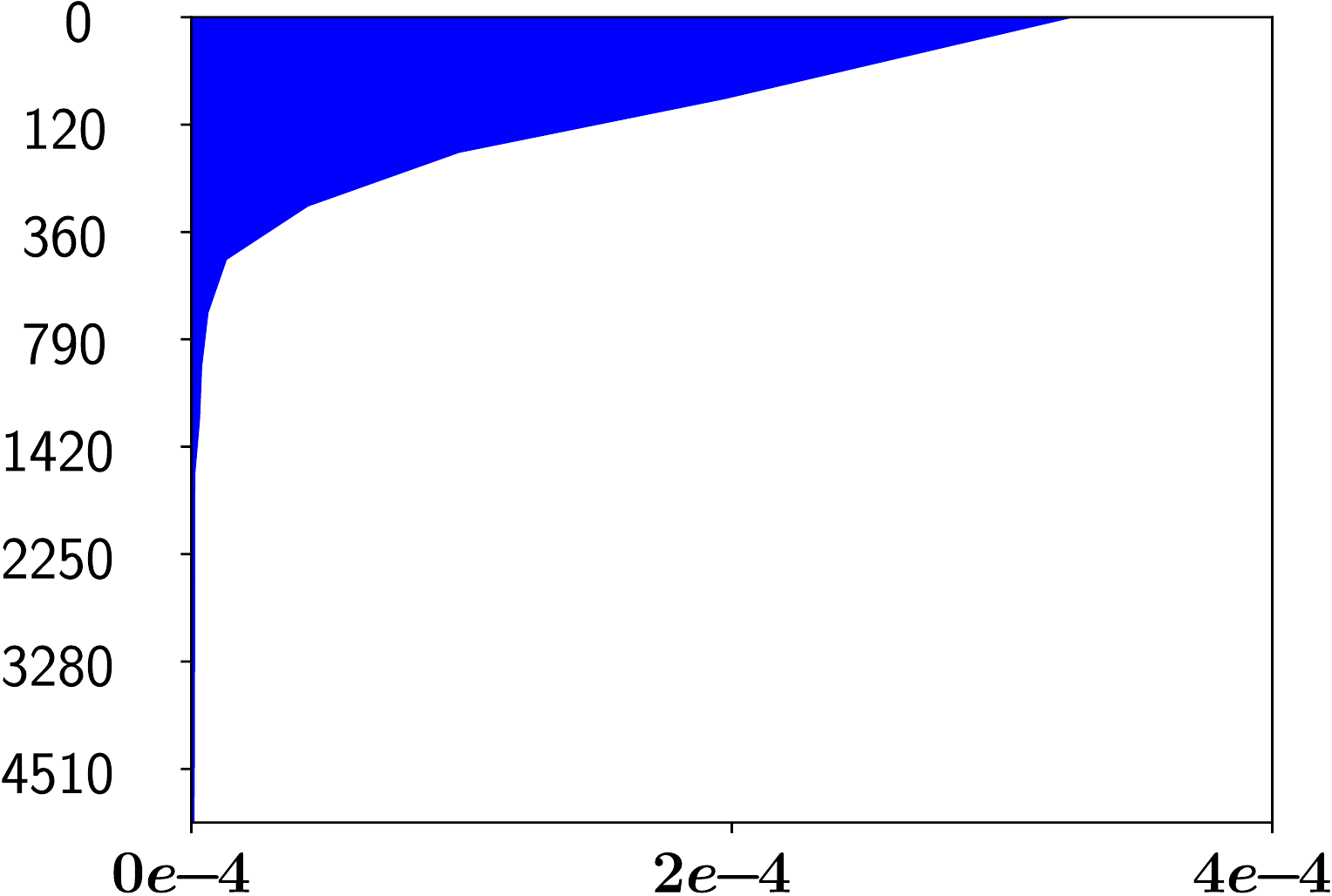}
       		\caption{average monthly change}
       		\label{fig:model_confidence:dop:time_diff}
       	\end{subfigure}
    \end{minipage}
    \hfill
    \begin{minipage}{0.49\textwidth}
       	\begin{subfigure}{1\linewidth}
       		\includegraphics[width=1.0\linewidth]{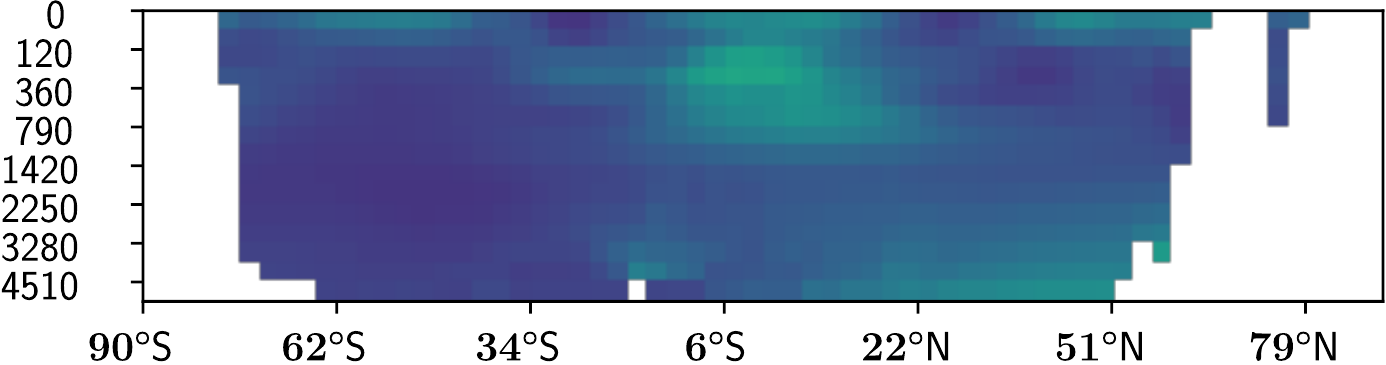}
       		\caption{Pacific Ocean: averaged over time and between 125$\degree$E and 70$\degree$W}
       		\label{fig:model_confidence:dop:pacific}
       	\end{subfigure}
        \par\smallskip
       	\begin{subfigure}{1\linewidth}
       		\includegraphics[width=1.0\linewidth]{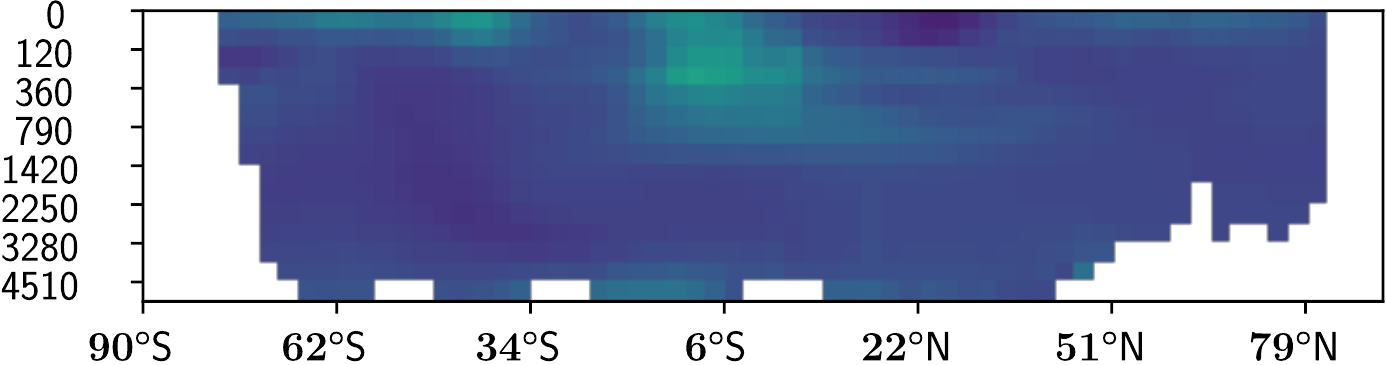}
       		\caption{Atlantic Ocean: averaged over time and between 70$\degree$W and 20$\degree$E} 
       		\label{fig:model_confidence:dop:atlantic}
       	\end{subfigure}
        \par\smallskip
       	\begin{subfigure}{1\linewidth}
       		\includegraphics[width=1.0\linewidth]{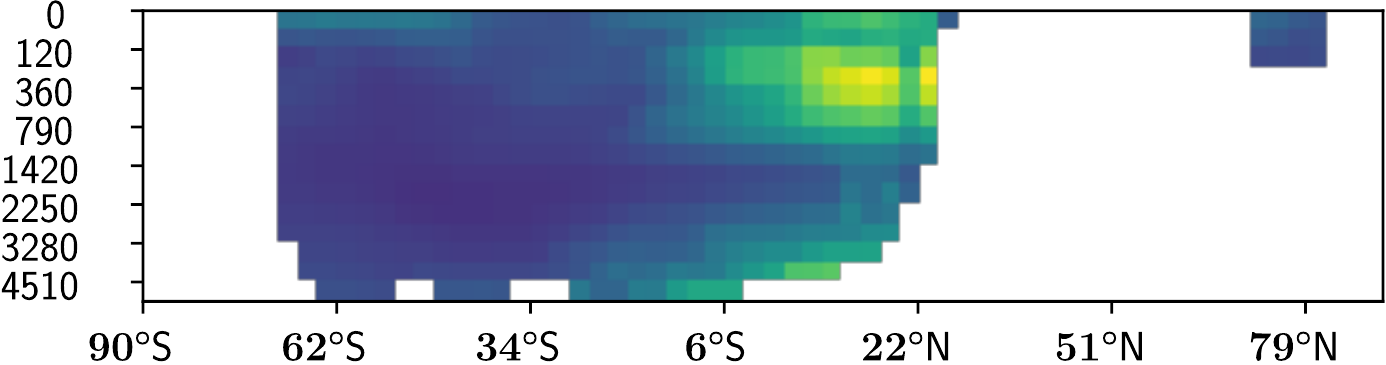}
       		\caption{Indian Ocean: averaged over time and between 20$\degree$W and 125$\degree$E}
       		\label{fig:model_confidence:dop:indian}
       	\end{subfigure}
    \end{minipage}
   	\caption{Confidence intervals length for dissolved organic phosphorus model output (in ${\rm mmol\,m}^{-3}$) with 99 \% confidence level.}
	\label{fig:model_confidence:dop} 
\end{figure}

The uncertainty in the model parameters implies uncertainty in the model output. This has been quantified as described in Subsection \ref{subsec: model uncertainty}. For each model output the uncertainty is quantified by the length of corresponding confidence intervals with confidence level of approximately 99 \%. Their lengths are plotted in Figure \ref{fig:model_confidence:po4} and \ref{fig:model_confidence:dop}.

The average uncertainty at the water surface is $6\times10^{-3}~{\rm mmol\,m}^{-3}$ for phosphate and $5\times10^{-3}~{\rm mmol\,m}^{-3}$ for dissolved organic phosphorus. This corresponds to an uncertainty relative to the average model output of around 1 \% for phosphate and around 2 \% for dissolved organic phosphorus. The uncertainty at the surface is high for both tracers right there where the dissolved organic phosphorus concentration itself is high.

With growing depth, the average uncertainty for phosphate decreases strictly monotonically. It is close to zero after roughly 700 m. In contrast, the uncertainty for dissolved organic phosphorus is almost constant over all depths.

The uncertainties near the surface change on average by 7\% per month for both tracers. The temporal variations regarding the uncertainties decrease with growing depth. There is almost no change over time deeper than 700 m for phosphate and deeper than 450 m for dissolved organic phosphorus. This corresponds to the model output itself which is almost constant over time from these depths on.

\subsection{Uncertainty Reduction by Additional Measurements}

The uncertainty regarding the model parameters as well as the model outputs can be reduced by additional measurements as described in Subsection \ref{subsec: uncertainty reduction}.

\begin{figure}[t]
    \begin{minipage}{0.50\textwidth}
       	\begin{subfigure}{1\linewidth}
       		\includegraphics[width=1.0\linewidth]{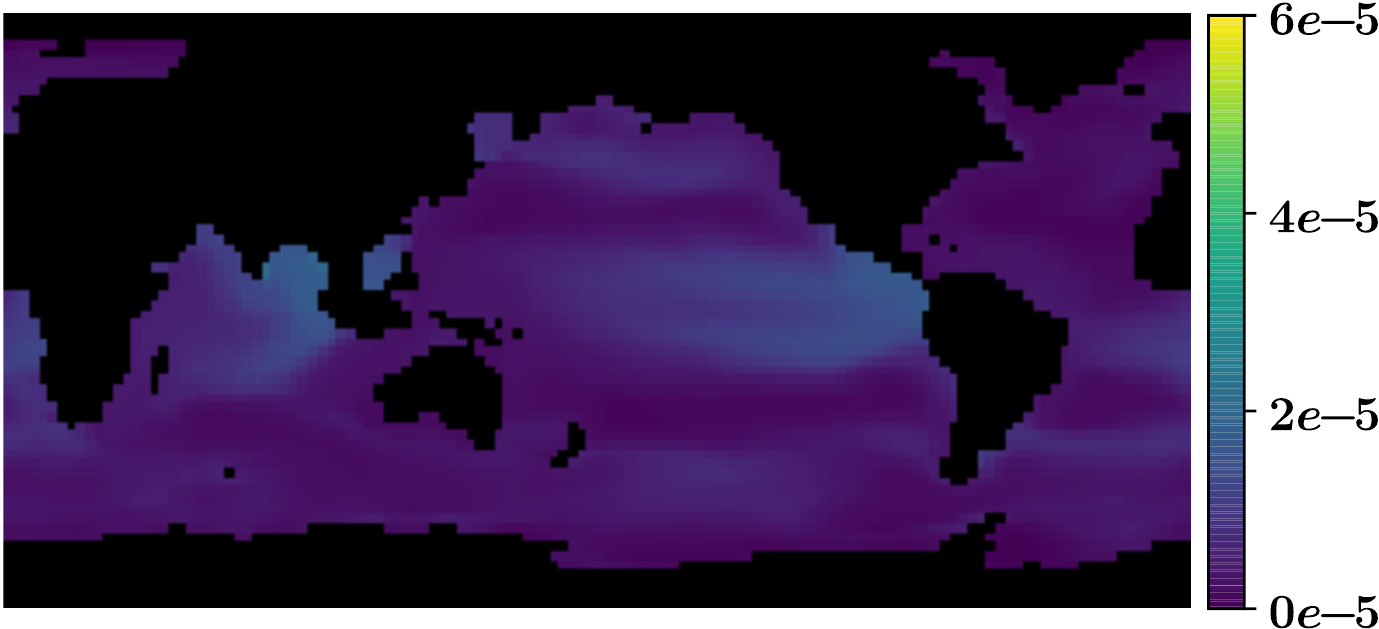}
          		\caption{water surface: averaged over time and 0 to 25 ${\rm m}$ depth} 
       		\label{fig:model_confidence_increase:po4:surface}
       	\end{subfigure}
        \par\smallskip
    	\begin{subfigure}{0.49\linewidth}
    		\includegraphics[width=1.0\linewidth, height=0.8\linewidth]{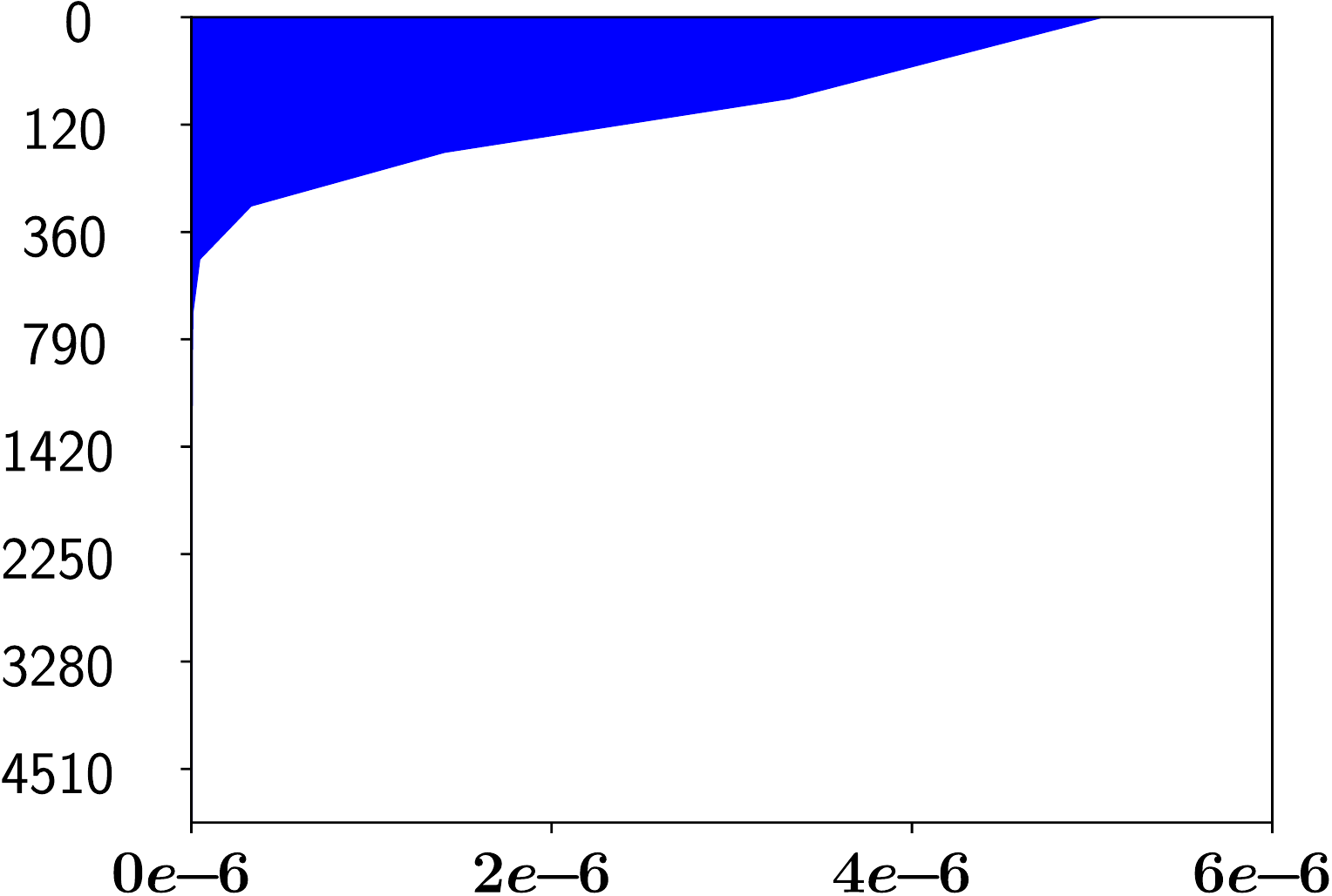}
    		\caption{averaged over all but depth} 
    		\label{fig:model_confidence_increase:po4:depth}
    	\end{subfigure}
       	\hfill
       	\begin{subfigure}{0.49\linewidth}
       		\includegraphics[width=1.0\linewidth, height=0.8\linewidth]{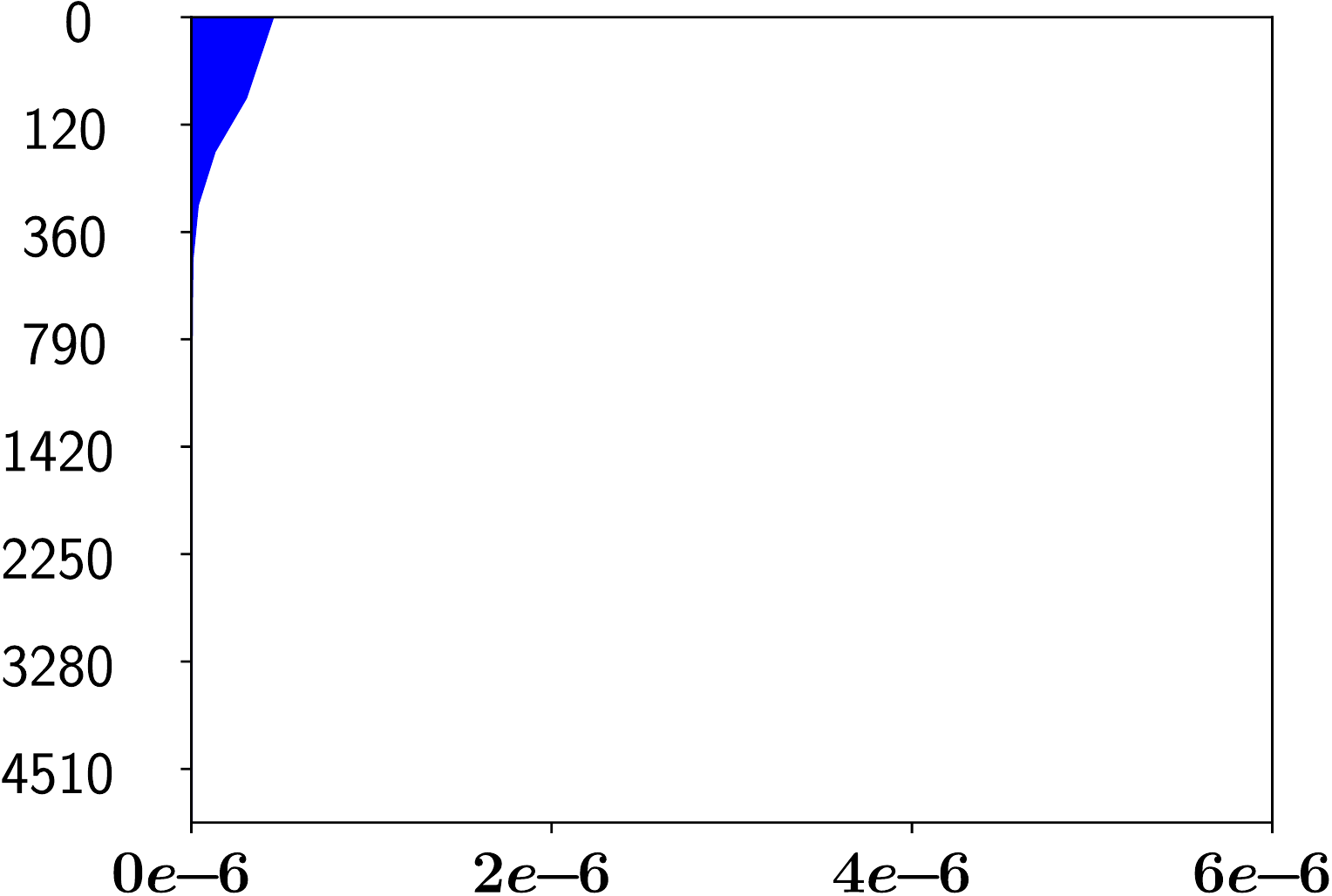}
            \caption{average monthly change}
       		\label{fig:model_confidence_increase:po4:time_diff}
       	\end{subfigure}
    \end{minipage}
    \hfill
    \begin{minipage}{0.49\textwidth}
       	\begin{subfigure}{1\linewidth}
       		\includegraphics[width=1.0\linewidth]{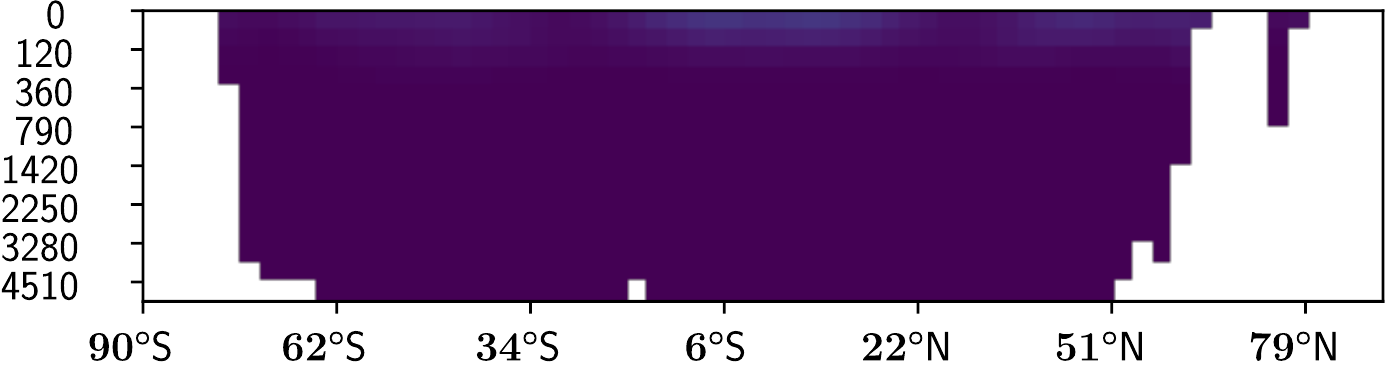}
       		\caption{Pacific Ocean: averaged over time and between 125$\degree$E and 70$\degree$W}
       		\label{fig:model_confidence_increasedence:po4:pacific}
       	\end{subfigure}
        \par\smallskip
       	\begin{subfigure}{1\linewidth}
       		\includegraphics[width=1.0\linewidth]{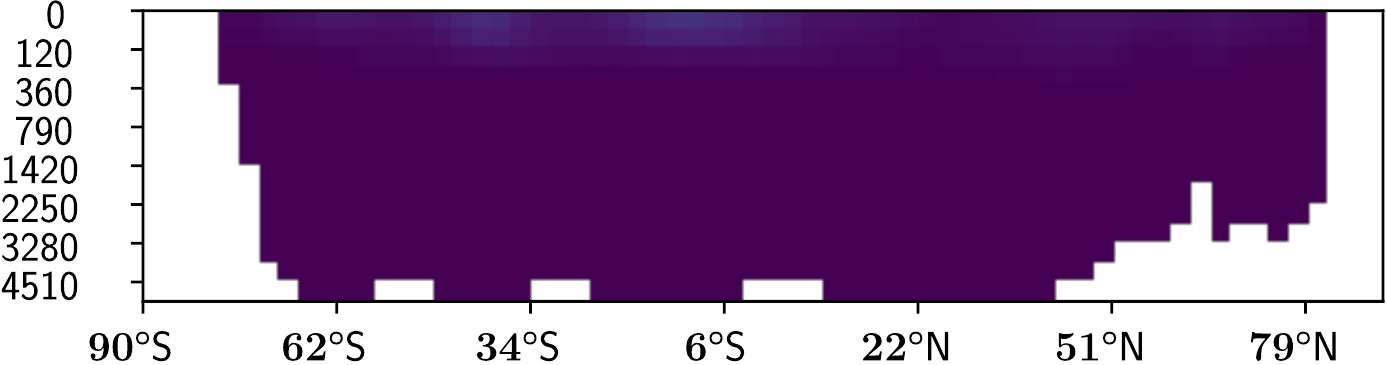}
       		\caption{Atlantic Ocean: averaged over time and between 70$\degree$W and 20$\degree$E} 
       		\label{fig:model_confidence_increase:po4:atlantic}
       	\end{subfigure}
        \par\smallskip
       	\begin{subfigure}{1\linewidth}
       		\includegraphics[width=1.0\linewidth]{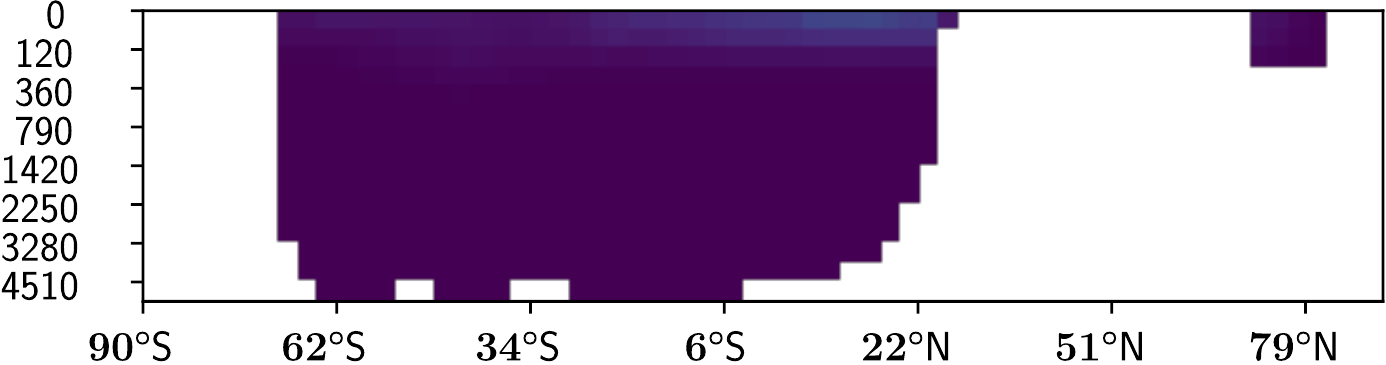}
       		\caption{Indian Ocean: averaged over time and between 20$\degree$W and 125$\degree$E}
       		\label{fig:model_confidence_increase:po4:indian}
       	\end{subfigure}
    \end{minipage}
	\caption{Uncertainty reduction by one phosphate measurement at this location (in ${\rm mmol\,m}^{-3}$).}
	\label{fig:model_confidence_increase:po4} 
\end{figure}

\begin{figure}[t]
    \begin{minipage}{0.50\textwidth}
    	\begin{subfigure}{1\linewidth}
    		\includegraphics[width=1.0\linewidth]{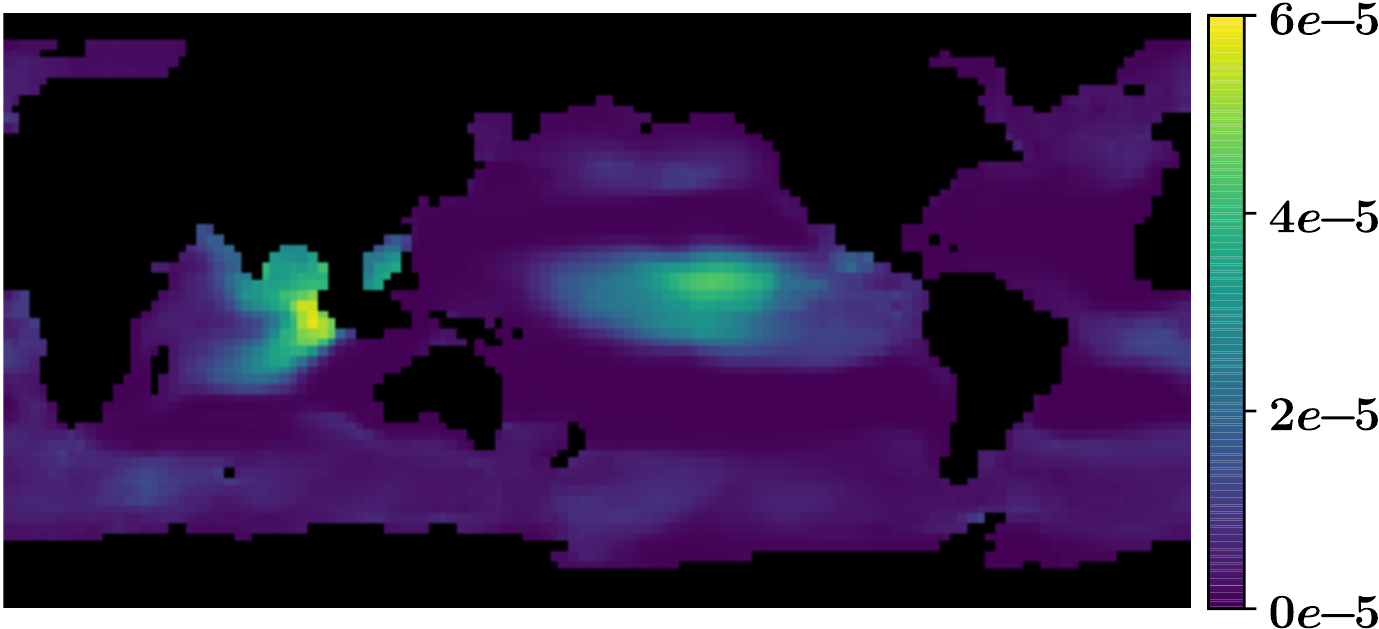}
       		\caption{water surface: averaged over time and 0 to 25 ${\rm m}$ depth} 
    		\label{fig:model_confidence_increase:dop:surface}
    	\end{subfigure}
        \par\smallskip
    	\begin{subfigure}{0.49\linewidth}
      		\includegraphics[width=1.0\linewidth, height=0.8\linewidth]{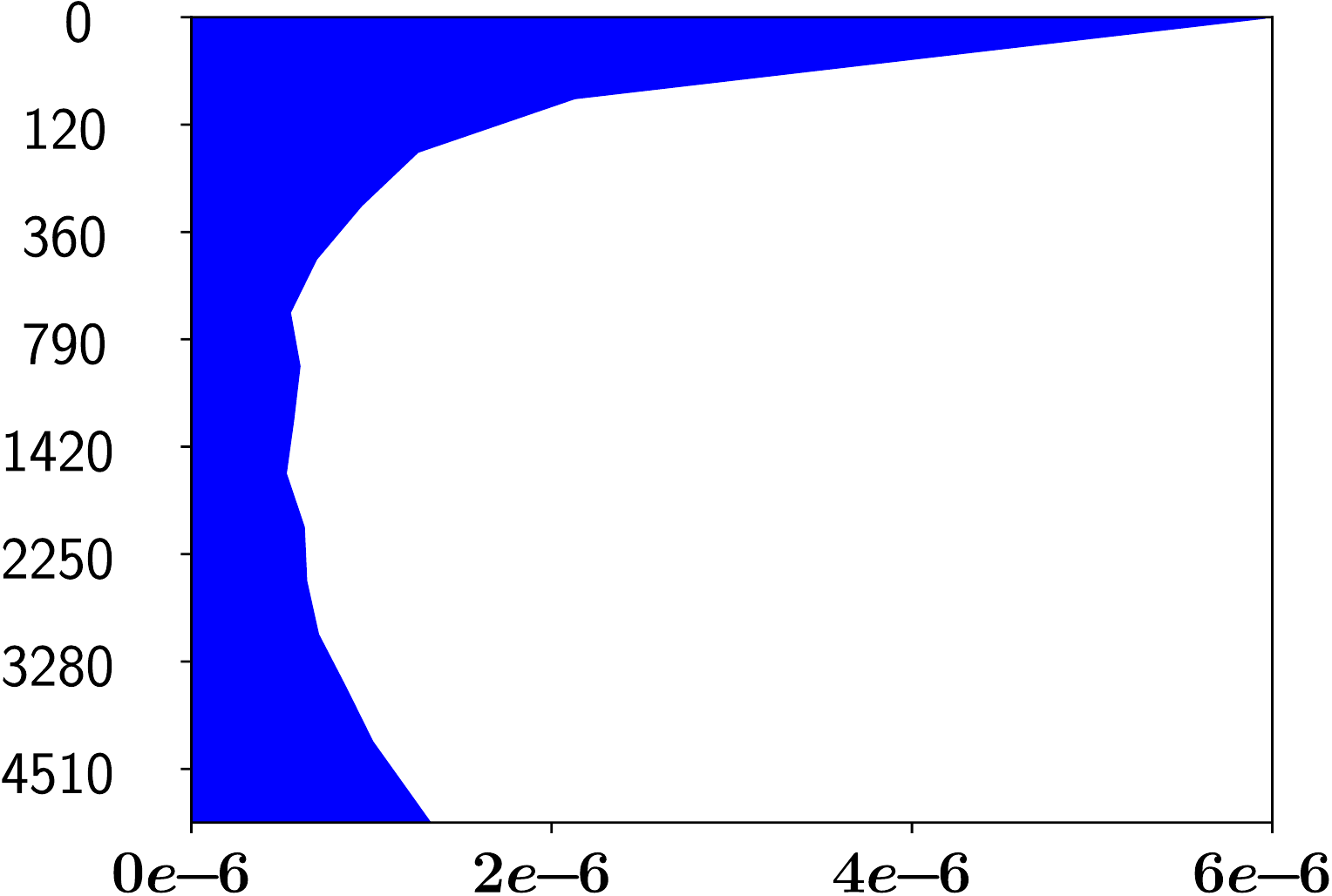}
    		\caption{averaged over all but depth} 
    		\label{fig:model_confidence_increase:dop:depth}
    	\end{subfigure}
       	\hfill
       	\begin{subfigure}{0.49\linewidth}
       		\includegraphics[width=1.0\linewidth, height=0.8\linewidth]{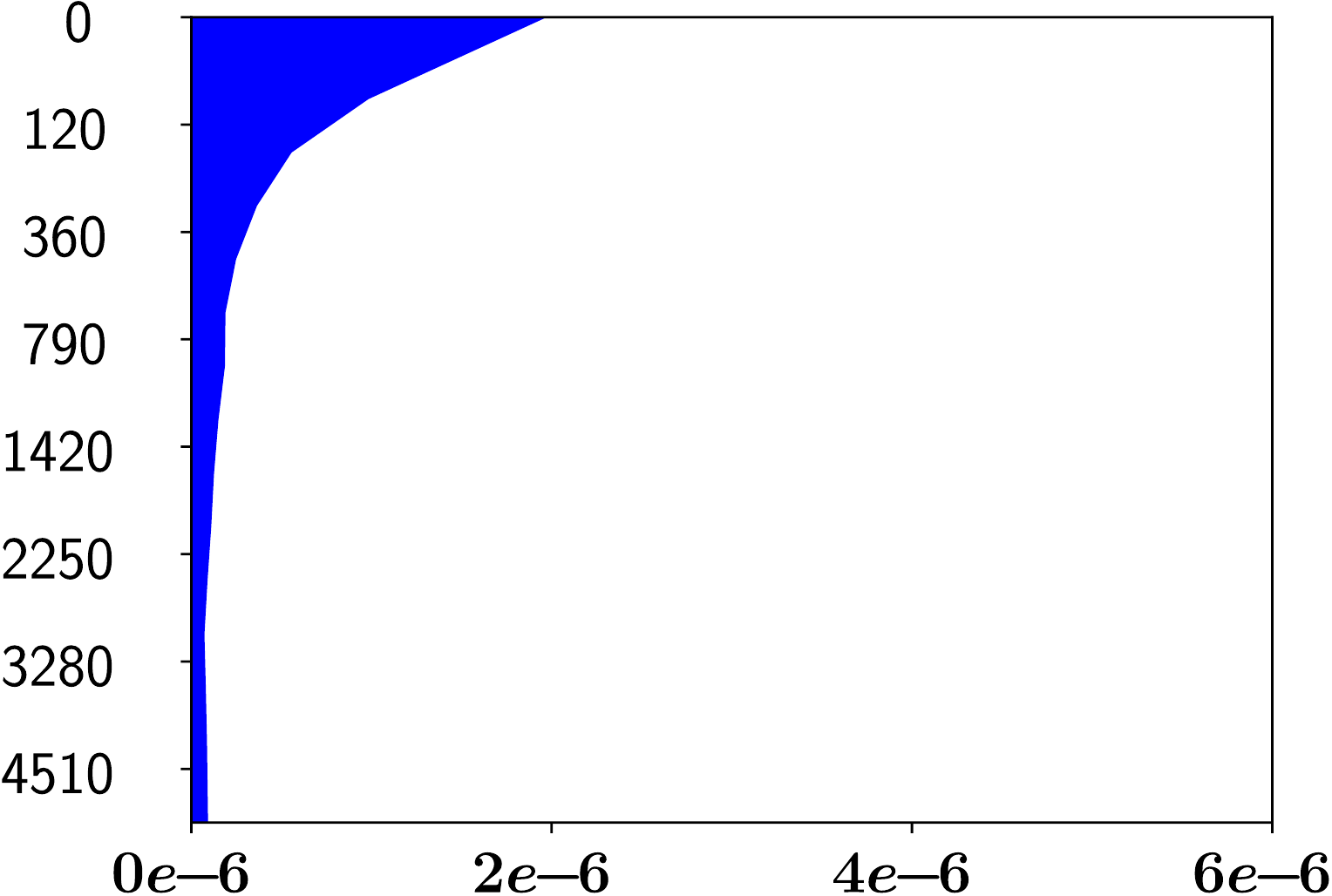}
            \caption{average monthly change}
       		\label{fig:model_confidence_increase:dop:time_diff}
       	\end{subfigure}
    \end{minipage}
    \hfill
    \begin{minipage}{0.49\textwidth}
       	\begin{subfigure}{1\linewidth}
       		\includegraphics[width=1.0\linewidth]{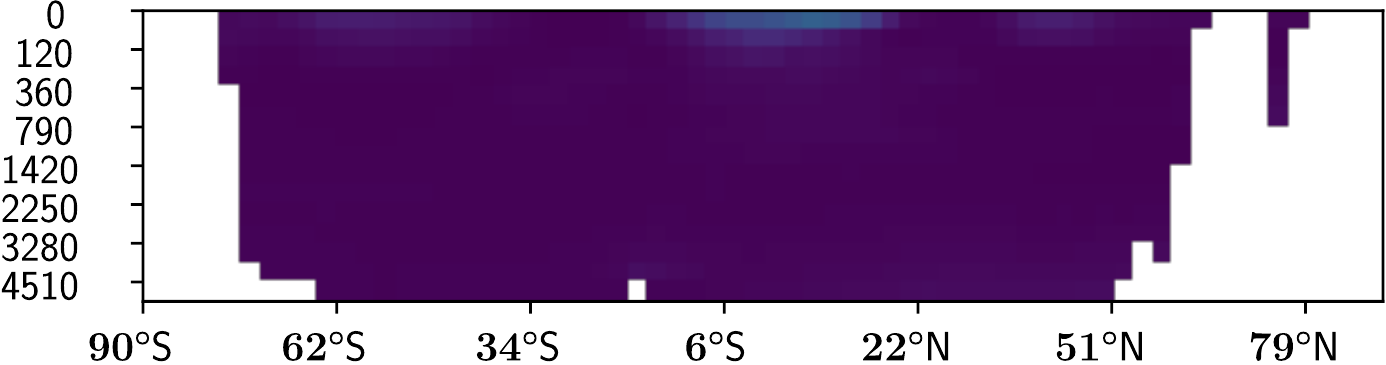}
       		\caption{Pacific Ocean: averaged over time and between 125$\degree$E and 70$\degree$W}
       		\label{fig:model_confidence_increase:dop:pacific}
       	\end{subfigure}
        \par\smallskip
       	\begin{subfigure}{1\linewidth}
       		\includegraphics[width=1.0\linewidth]{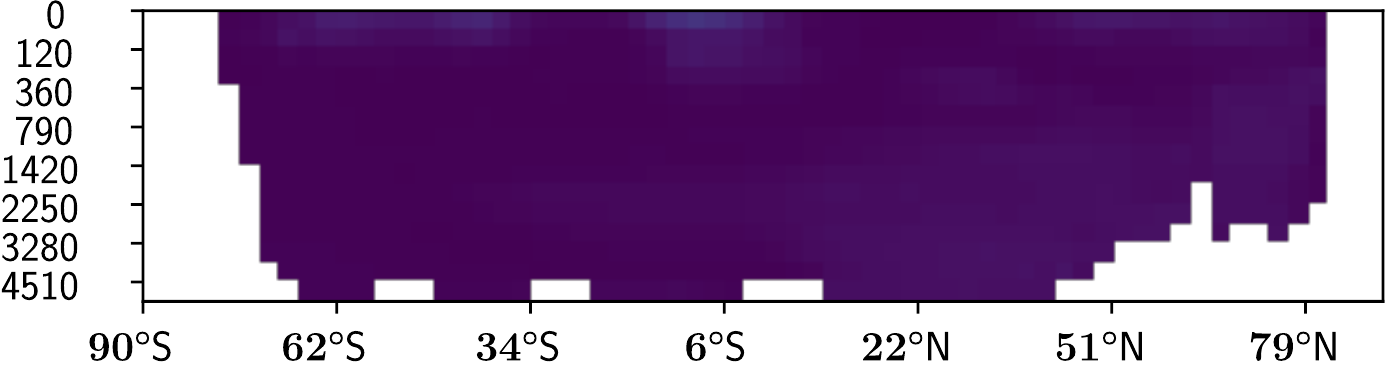}
       		\caption{Atlantic Ocean: averaged over time and between 70$\degree$W and 20$\degree$E} 
       		\label{fig:model_confidence_increase:dop:atlantic}
       	\end{subfigure}
        \par\smallskip
       	\begin{subfigure}{1\linewidth}
       		\includegraphics[width=1.0\linewidth]{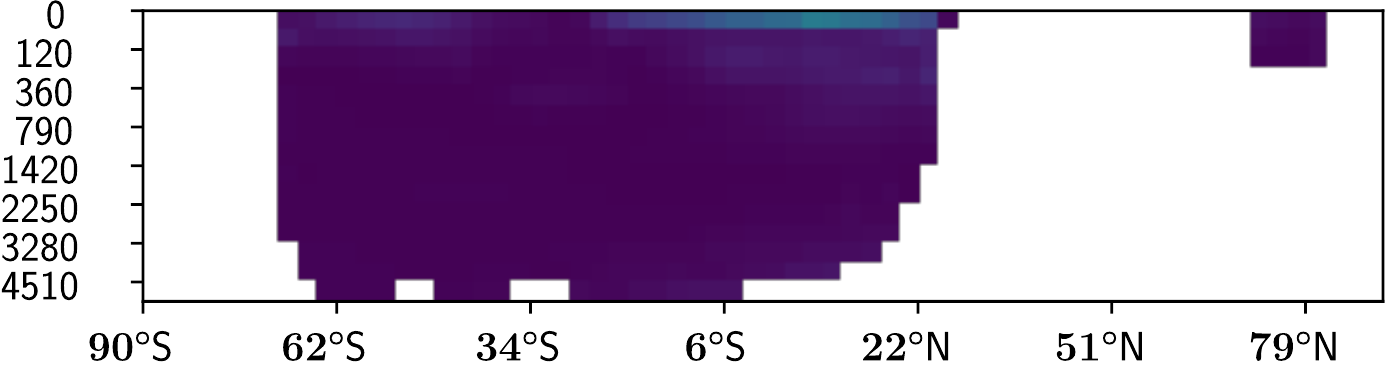}
       		\caption{Indian Ocean: averaged over time and between 20$\degree$W and 125$\degree$E}
       		\label{fig:model_confidence_increase:dop:indian}
       	\end{subfigure}
    \end{minipage}
  	\caption{Uncertainty reduction by one dissolved organic phosphorus measurement at this location (in ${\rm mmol\,m}^{-3}$).}
	\label{fig:model_confidence_increase:dop} 
\end{figure}

In order to find out which measurement design significantly reduce the uncertainties and which result only in a slight information gain, we have analyzed the average model uncertainty equally weighted for both tracers as described in Equation \eqref{eq: oed criterion: relative average model output uncertainty} resulting for one additional measurement. Figure \ref{fig:model_confidence_increase:po4} and \ref{fig:model_confidence_increase:dop} show by what proportion the average model uncertainty is reduced by one additional measurement at this point.

The most informative measurements are located at the water surface. The information content decreases rapidly with growing depth. Compared to the the information content at the surface, it is below one third for phosphate measurements deeper than 150 m and for dissolved organic phosphorus measurements deeper than 80 m. Deeper than 400 m it is close to zero for phosphate measurements and deeper than 200 m it is approximately constant one sixth for dissolved organic phosphorus measurements. The time of the measurement seems to have little effect on their information content.

Phosphate measurements have the highest information content at the north-eastern part of the Indian Ocean and at the middle-east part of the Pacific Ocean. This indicates that measurements in areas where the concentration of dissolved organic phosphorus is high are especially worthwhile for phosphate measurements.

The highest information content of dissolved organic phosphorus measurements is at the surface of the north-eastern part of the Indian Ocean and the middle of the Pacific Ocean both around the equator. This indicates that measurements in areas where the concentration of dissolved organic phosphorus is high but that of phosphorus is low are especially worthwhile for dissolved organic phosphorus measurements.

A dissolved organic phosphorus measurement contains usually twice as much information as a phosphate measurements. However, carrying out a dissolved organic phosphorus measurement is many times more complex and expensive than carrying out a phosphate measurement. This means that carrying out dissolved organic phosphorus measurements is not worthwhile for reducing the model uncertainty.

A single additional measurement can reduce the average model uncertainty at most by roughly a twenty thousandth part. Hence, for a significant reduction, many additional measurements are required. This is plausible, since more than four million measurements have already been carried out and result in the current uncertainty.

 \conclusions  \label{sec: conclusion}

In this article we have presented several methods for model parameter estimation and uncertainty quantification. They are based on the generalized least squares estimator which has been described together with its statistical properties.

Several approximations of the covariance matrix of the estimator of the model parameters as well as the corresponding model output have been introduced. They are based on the first and second derivative of the model regarding its parameters. Their advantages and disadvantages have been emphasized. Approximate confidence intervals were provided as another way to quantify uncertainties.

Optimal experimental design methods have been briefly introduced which allow to predict the uncertainty reduction by additional measurements and to design new measurements in such a way that the information gain is maximized.

We have applied all these methods to a model for phosphate and dissolved organic phosphorus concentrations in the global ocean. For this, we have introduced the model briefly as well as its evaluation and corresponding measurement data.

We were able to find model parameters which are significantly more consistent with the measurement data compared to our initial guess. The individual model parameters of the model are subject to very diverse uncertainties. The uncertainties vary from 0.1 \% to 7 \% of the parameter values.

The uncertainties in the associated model output vary greatly as well, depending on location, time and tracer. The largest uncertainties are at the water surface, where, they are in average around 1 \% of the phosphate concentrations and around 2 \% of the dissolved organic phosphorus concentrations. Usually, they are high where the dissolved organic phosphorus concentration is high. With increasing depth the uncertainty for phosphate decreases rapidly while remaining more or less constant for dissolved organic phosphorus.

New measurements are most informative if they are close to the water surface. Phosphate measurements are especially worthwhile where the concentration of dissolved organic phosphorus is high. Taking into account the additional effort and costs associated with dissolved organic phosphorus measurements they are not worthwhile. If dissolved organic phosphorus measurements should be carried out nevertheless, they should be carried out where the dissolved organic phosphorus concentration is high and the phosphorus concentration is low. 

The results obtained for this model help to better assess its parameters and output as well as to plan new measurements. The applicability and usefulness of the presented methods has been shown with this application example.

\bibliographystyle{copernicus}
\bibliography{article}

\end{document}